\documentclass[10pt,conference]{IEEEtran}
\usepackage{cite}
\usepackage{amsmath,amssymb,amsfonts}
\usepackage{algorithmic}
\usepackage{graphicx}
\usepackage{textcomp}
\usepackage{xcolor}
\usepackage[hyphens]{url}
\usepackage{fancyhdr}
\usepackage{hyperref}
\hypersetup{
  pdftitle={BS-Spec: BFP-Based Bit-Sharing for Self-Speculative LLM Decoding with LUT-Driven Mixed-Precision Acceleration},
  pdfauthor={Jiwoo Kim, Gunho Park, Sunwoo Yoo, Baeseong Park, Dongsoo Lee, Youngjoo Lee},
  pdfsubject={LLM Inference Acceleration, Speculative Decoding, Computer Architecture}
}

\usepackage{booktabs}
\usepackage{tabularx}
\usepackage{makecell}
\usepackage{multirow} 

\newcommand{\hpcayear}{2027}

\title{BS-Spec: BFP-Based Bit-Sharing for Self-Speculative LLM Decoding with LUT-Driven Mixed-Precision Acceleration}

\def\hpcacameraready{} 
\newcommand\hpcaauthors{Jiwoo Kim$\dagger$, Gunho Park$\ddagger$, Sunwoo Yoo$\S$, Baeseong Park$\ddagger$, Dongsoo Lee $\ddagger$, and Youngjoo Lee$\dagger$}
\newcommand\hpcaaffiliation{KAIST$\dagger$, a2sys$\ddagger$, POSTECH$\S$}
\newcommand\hpcaemail{jiwoo.kim@kaist.ac.kr, youngjoo@kaist.ac.kr} 

\author{
  \ifdefined\hpcacameraready
    \IEEEauthorblockN{\hpcaauthors{}}
      \IEEEauthorblockA{
        \hpcaaffiliation{} \\
        \hpcaemail{}
      }
  \else
    \IEEEauthorblockN{\normalsize{HPCA \hpcayear{} Submission
      \textbf{\#\hpcasubmissionnumber{}}} \\
      \IEEEauthorblockA{
        Confidential Draft \\
        Do NOT Distribute!!
      }
    }
  \fi 
}

\fancypagestyle{camerareadyfirstpage}{%
  \fancyhf{} 
  
}

\begin{document}
\maketitle

\ifdefined\hpcacameraready 
  \thispagestyle{camerareadyfirstpage}
  \pagestyle{empty}
\else
  \thispagestyle{plain}
  \pagestyle{plain}
\fi

\newcommand{\hpcaheight}{0mm}
\ifdefined\eaopen
\renewcommand{\hpcaheight}{12mm}
\fi


\begin{abstract}
As large language models (LLMs) are increasingly deployed across diverse inference services, serving workloads span a wide range of batch sizes, context lengths, and performance requirements. Speculative decoding improves inference throughput, but conventional approaches require a separate draft model and KV cache, while existing self-speculative methods suffer from workload-insensitive draft configurations and inefficient mixed-precision execution.

In this paper, we present BS-Spec, a block floating point (BFP)-based bit-sharing framework that jointly optimizes draft construction, workload-aware precision selection, and hardware execution. BS-Spec derives multiple draft representations directly from a single BFP target model by partially loading the most significant bits of weights and the KV cache while preserving shared exponents, eliminating separate draft-model storage. Offline calibration selects the weight and KV-cache precision pair that maximizes estimated speedup for each serving regime, and a precision lookup table applies the selected configuration at runtime.
To support this adaptive execution, we design a mixed-precision BFP accelerator with a lookup table (LUT)-based variable-precision datapath for both draft and target computation. Experimental results show that BS-Spec achieves up to 1.9$\times$ speedup over conventional autoregressive decoding and improves speedup over the evaluated bit-sharing self-speculative decoding baselines by 67.2\% on average, demonstrating the benefits of jointly co-designing bit sharing, precision adaptation, and specialized hardware for diverse LLM serving workloads.

\end{abstract}

\section{Introduction}

Large language models (LLMs) have rapidly grown in both parameter count and context length to support increasingly capable inference \cite{openai2024gpt4technicalreport, grattafiori2024llama}. 
As a result, the autoregressive decoding bottleneck has shifted away from raw compute throughput toward memory movement \cite{sadhukhan2024magicdec}. 
Speculative decoding has emerged as a promising approach by replacing multiple expensive target-model decoding steps with lightweight draft generation followed by a single verification step.
By allowing a smaller draft model to propose multiple candidate tokens before verification, speculative decoding improves effective throughput while preserving the final output \cite{leviathan2023fast, kim2023speculative, chen2023accelerating}.
However, to realize these gains, the costs of draft-generation need to be well balanced with the number of accepted tokens \cite{sadhukhan2024magicdec}.
Achieving this balance in practice is challenging.

\begin{figure*}[t]
    \centering
    \includegraphics[width=0.98\textwidth]{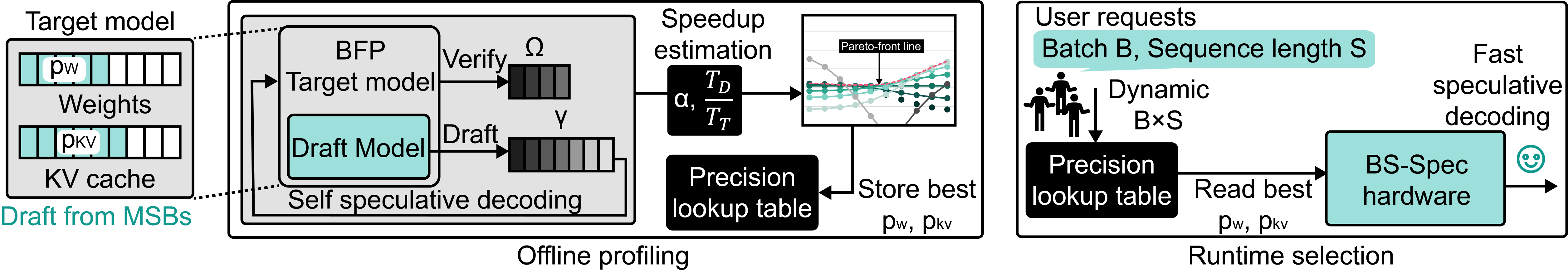}
    \caption{\textcolor{black}{Overview of BS-Spec software-hardware co-optimization flow.}}
    \label{fig:overall_concept}
\end{figure*}

In speculative decoding, maintaining a separate draft model can impose additional memory overhead from its weights and KV cache, which may constrain deployment under memory-limited serving conditions \cite{elhoushi2024layerskip, tiwari2025quantspec}.
Self-speculative decoding addresses this issue by deriving draft predictions directly from the target model itself \cite{liu2024kangaroo}. 
Its effectiveness depends on maintaining strong alignment with the target model to preserve high acceptance rates while keeping draft generation substantially cheaper than full-precision decoding.
Existing approaches, however, often struggle to balance draft efficiency and accuracy. Layer-skipping methods are constrained by the limited number of layers that can be skipped without substantially degrading draft quality \cite{elhoushi2024layerskip, zhang2024draft}. Intermediate feature-reuse methods \cite{cai2024medusa, ankner2024hydra, li2024eagle} reduce drafting overhead but typically require additional training; moreover, learned drafters such as EAGLE\cite{li2024eagle} may exhibit sharply reduced acceptance lengths under long-context decoding and out-of-distribution \cite{lee2025owl}.
Quantization-based self-speculative decoding approaches are particularly promising in this regard. Quantized LLMs have been observed to better preserve the reasoning ability of the original model \cite{srivastava2025towards}, making them appealing candidates for draft generation. 
However, existing methods do not perform consistently across diverse inference scenarios because they optimize weight and KV-cache costs in isolation and rely on fixed draft configurations. These designs perform well only in limited workload regions and fail to maintain efficiency across the full serving space.

In this paper, we present BS-Spec, a block floating point (BFP)-based bit-sharing framework for self-speculative LLM decoding that jointly adapts weight and KV-cache precisions to maintain high efficiency across diverse serving scenarios. Fig.~\ref{fig:overall_concept} illustrates the overall concept of BS-Spec. BS-Spec constructs lower-precision draft representations by partially loading only the most significant bits (MSBs) of the target model weights and KV cache. Because BFP significands are already aligned under a shared exponent \cite{kim2023winning,noh2023flexblock}, this partial-loading process requires no additional conversion or reconstruction. Consequently, BS-Spec supports multiple draft precisions from a single target representation without separate draft-model storage or preprocessing. 

More specifically, BS-Spec determines the draft precision through offline calibration and stores the selected configurations in a precision lookup table (PLT).
At runtime, the BS-Spec controller queries the PLT according to the current serving workload and selects the corresponding optimal weight and KV-cache precisions. In memory-bound LLM inference, lower precision reduces memory-access cost but may decrease the acceptance rate, whereas higher precision improves draft fidelity at a greater execution cost. This trade-off varies with the serving regime, particularly with batch size and sequence length \cite{tiwari2025quantspec}. Consequently, the optimal draft precision varies across workloads depending on their memory and compute characteristics. By jointly considering draft cost and acceptance rate, BS-Spec adapts the weight and KV-cache precisions to maximize speculative-decoding speedup. Efficient execution of BS-Spec further requires a hardware engine capable of processing BF16 activations with BFP-formatted weights and KV cache while dynamically varying the significand precision from BFP5 to BFP9.

To this end, we design a bit-serial mixed-precision datapath that operates directly on the sign-magnitude BFP representation, enabling low-overhead precision scaling without conventional floating-point-centric accumulation. The design combines LUT-assisted accumulation, which reduces the cycle overhead of bit-serial execution, with a LUT-stationary SIMD organization that allows multiple weight lanes within a compute unit to reuse the same LUT entries. With these optimizations, BS-Spec achieves the highest speedup over 16-bit autoregressive decoding and the lowest energy consumption among the state-of-the-art quantization-based self-speculative decoding methods.

In summary, this paper makes the following contributions:
\begin{itemize}
\item \textbf{BFP-based bit-sharing self-speculative decoding.}
    We propose BS-Spec, which derives multiple draft precisions from a single BFP target representation through MSB-only partial loading under shared exponents, eliminating separate draft-model storage and additional conversion.

\item \textbf{Theoretical workload-aware precision optimization}
    We develop an offline calibration scheme that jointly selects weight and KV-cache precisions based on draft cost and acceptance rate, and stores the optimal configurations in a PLT for runtime lookup.

\item \textbf{Hardware support for bit-shared BFP execution.}
    We develop a bit-serial accelerator that directly executes runtime-variable BFP5–BFP9 weights and KV cache with BF16 activations. Its BFP-aware LUT generator folds per-group exponent alignment into LUT construction, while a decoding-oriented dataflow efficiently processes the selected bit planes for both draft and target execution.
\end{itemize}

\begin{table}[t]
\caption{Arithmetic intensity of representative linear and attention operations during autoregressive decoding. $Batch$ is batch size, $T$ is sequence length, $D_M$ is model dimension, $D_H$ is head dimension, $N_H$ is the number of attention heads, and $N_{KV}$ is the number of KV heads.}
\begin{tabularx}{0.48\textwidth}{c|ccc}
\toprule[1.2pt]
                       & Linear                  &   & Attention  \\ 
\midrule
MOP                    & $O({D_M}^2)$            &   & $O(D_H \cdot T)$         \\
FLOPs                  & $O(Batch \cdot {D_M}^2)$    &   & $O(\frac{N_H}{N_{KV}} \cdot T \cdot D_H)$ \\
Arithmetic   Intensity & $O(Batch)$                  &  & $O(\frac{N_H}{N_{KV}})$ \\
\bottomrule[1.2pt]

\end{tabularx}

\label{tab:arithmetic_intensity}
\end{table}

\begin{table}[]

\caption{\textcolor{black}{Comparison of autoregressive decoding (AR), conventional speculative decoding (SD), and self-speculative decoding (Self-SD).}}

\begin{tabularx}{0.5\textwidth}{c|ccc}
\toprule[1.2pt]
    & AR &SD & Self-SD             
    \\ \midrule
\makecell[c]{Resident models}        & Target only                                 & Target + draft                      & \textbf{Target only}
\\

\makecell[c]{Draft phase}            & N/A                                           & Draft model             & \textbf{Target subset}
\\
\makecell[c]{Verification phase}      & Target model                    & Target model            & \textbf{Target model}
\\

\makecell[c]{Acceleration}            & No                                      & Yes                                 & \textbf{Yes}
\\

\makecell[c]{Additional memory} & No                                        & Required                            & \textbf{No}
\\ 
\bottomrule[1.2pt]
\end{tabularx}{}
\label{tab:ar_sd_ssd}
\end{table}

\begin{figure}[t]
    \centering
    \includegraphics[width=0.49\textwidth]{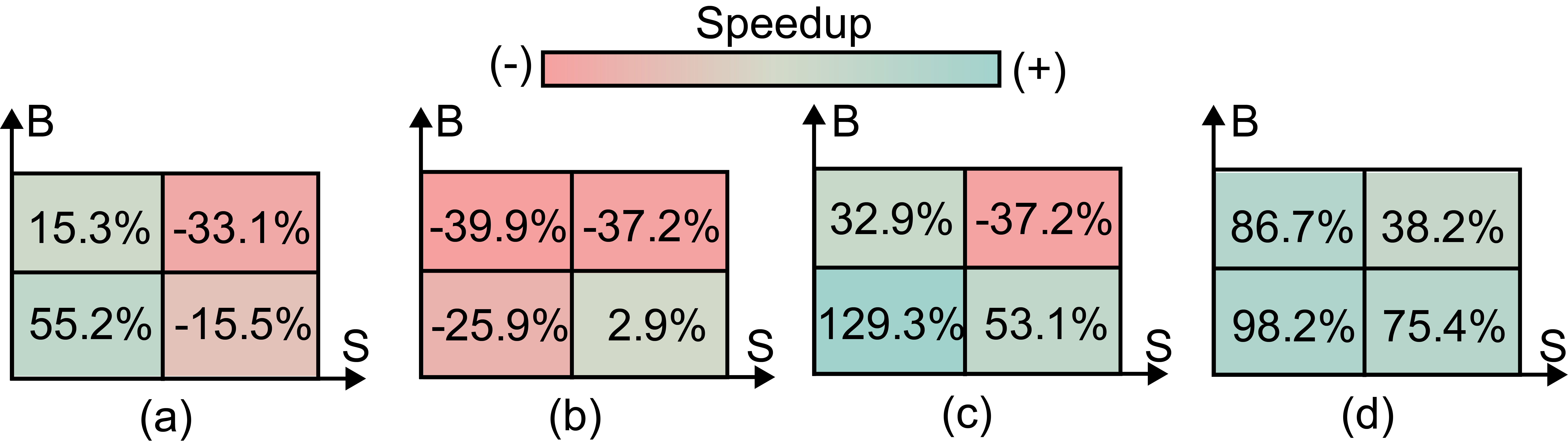}
    \caption{Speculative decoding speedup across varying sequence lengths ($S$) and batch sizes ($B$) for (a) SPEQ\cite{zhao2025quarter}, QuantSpec \cite{tiwari2025quantspec} with (b) fully self-speculative variant, (c) separately stored quantized weights, and (d) the proposed BS-Spec with joint weight--KV-cache precision adaptation.}
    \label{fig:b_s_efficiency}
\end{figure}

\begin{figure}[t]
    \centering
    \includegraphics[width=0.5\textwidth]{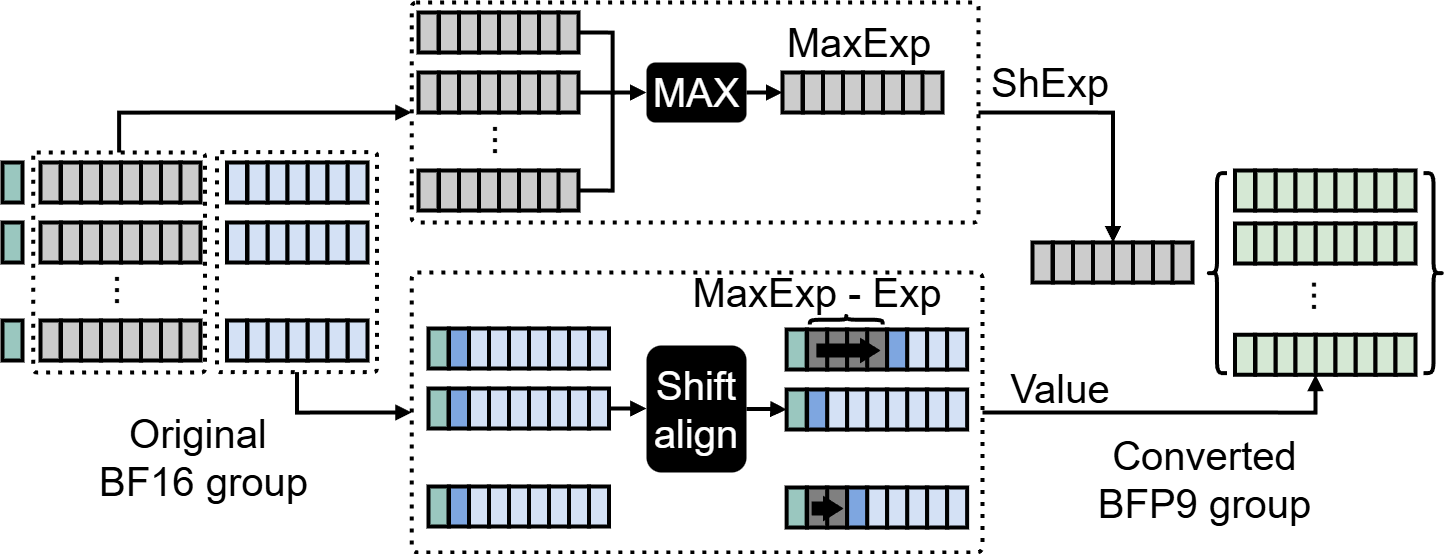}
    \caption{\textcolor{black}{Procedure to generate BFP format with per-group shared exponent.}}
    \label{fig:bfp_make}
\end{figure}

\begin{figure}[t]
    \centering
    \includegraphics[width=0.5\textwidth]{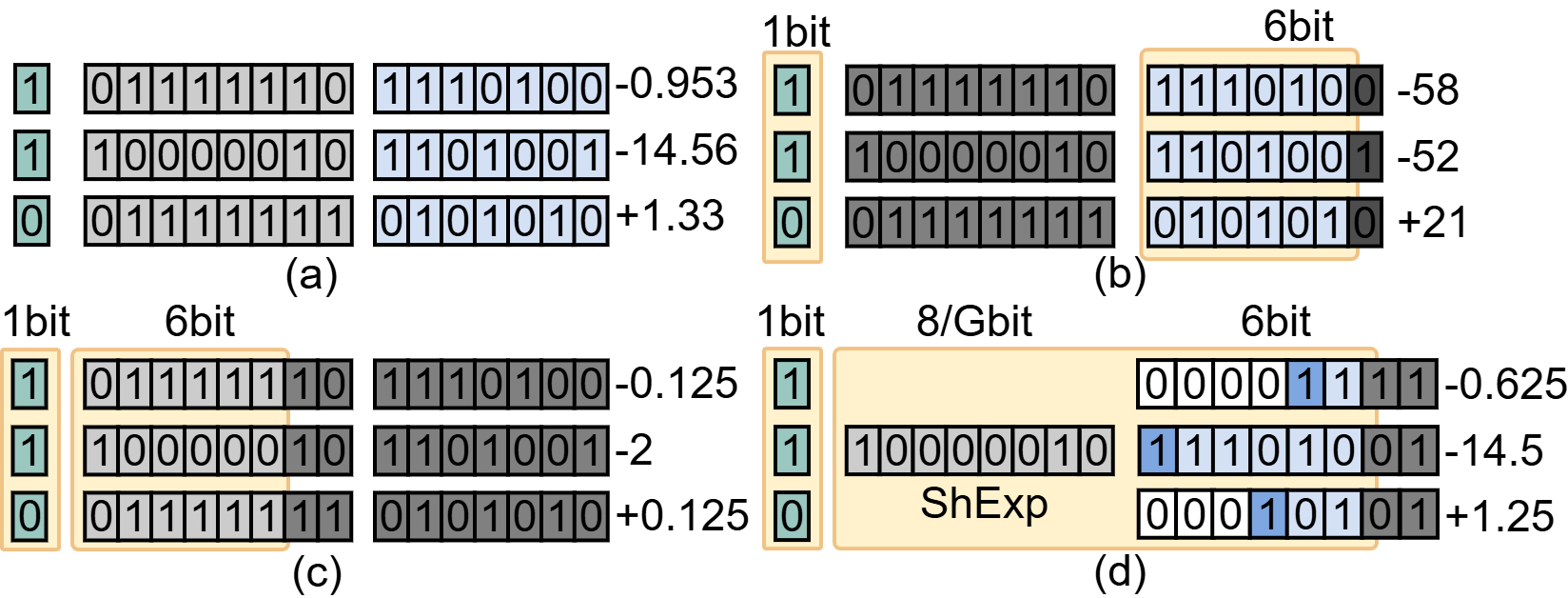}
    \caption{\textcolor{black}{Numerical example of (a) the original BF16 representation and 7-bit partial loads, each comprising one sign bit and 6-bit from (b) the BF16 mantissa, (c) the BF16 exponent (SPEQ \cite{zhao2025quarter}-style), and (d) the aligned BFP9 significand in the proposed scheme.}}
    \label{fig:numerical_example}
\end{figure}

\begin{figure*}[t]
    \centering
    \includegraphics[width=1\textwidth]{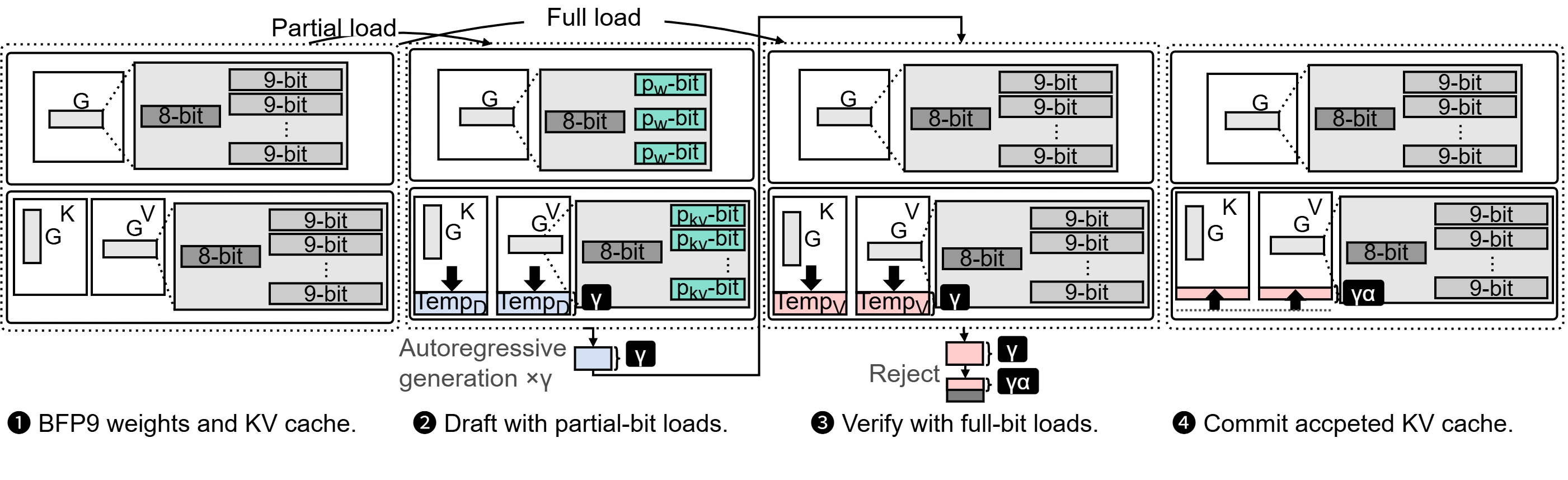}
    \caption{\textcolor{black}{Detailed workflow of BS-Spec: draft tokens are generated using partially loaded $p_w$-bit weights and $p_{kv}$-bit KV cache, then verified using the fully loaded weights and KV cache. The KV cache is subsequently updated and retained at full precision.}}
    \label{fig:overview}
\end{figure*}

\section{Background}

\subsection{Memory Bottlenecks in Autoregressive LLM Decoding}

Decoder-only LLM inference consists of two phases: parallel prefill and autoregressive decoding \cite{patel2024splitwise}. During prefill, all input tokens are processed in parallel to build the KV cache and generate the first output token. In contrast, autoregressive decoding generates tokens sequentially, repeatedly invoking the model while attending to the growing KV cache. As a result, the two phases exhibit different performance characteristics: prefill is dominated by large batched matrix multiplications, whereas decoding is dominated by repeated accesses to model weights and the KV cache \cite{zhong2024distserve}.

Autoregressive decoding remains a critical phase in LLM serving because it generates tokens sequentially while repeatedly accessing model weights and the growing KV cache. Its bottleneck can be characterized by arithmetic intensity, defined as the number of floating-point operations performed per byte of memory traffic \cite{williams2009roofline}. As summarized in Table~\ref{tab:arithmetic_intensity}, the arithmetic intensity of linear layers increases with batch size but remains low under practical serving configurations, making weight loading predominantly memory-bound.

At larger batch sizes and longer contexts, attention and KV-cache accesses account for an increasing fraction of execution cost. Although modern LLMs adopt MHA, GQA, or MQA to improve KV reuse \cite{ainslie2023gqa}, attention still repeatedly accesses the growing KV cache and remains memory-bound in practical decoding workloads \cite{hooper2024kvquant}. Therefore, decoding is dominated by weight traffic in shorter-context regimes and by KV-cache traffic in longer-context regimes \cite{he2024zipcache, dong2024qaq}.
Because the relative costs of weight and KV-cache accesses vary substantially across serving workloads, optimizing either component in isolation inevitably limits the benefits to a narrow operating region; achieving consistently high efficiency requires jointly adapting both to the current workload \cite{tiwari2025quantspec}.

\subsection{Speculative Decoding and the Draft Cost-Acceptance Tradeoff}

Speculative decoding improves LLM inference by allowing a lightweight draft model to generate multiple candidate tokens before verification by the target model \cite{leviathan2023fast, chen2023accelerating}.
When drafted tokens are accepted, the target model can skip several decoding steps, improving effective throughput without changing the final output.
In practice, the benefit of speculative decoding depends on the balance between draft-generation cost and the number of accepted draft tokens\cite{sadhukhan2024magicdec}. This tradeoff can be described by the formulation:
\begin{equation}
\frac{T_{Avg}^{SD}}{T_T}=\frac{1}{\Omega (\gamma, \alpha)} \left( \frac{\gamma \cdot T_D}{T_T} + \frac{T_V(\gamma)}{T_T} \right)
\label{eq:sd_speedup}
\end{equation}
This expression shows that speculative-decoding speedup is governed by three factors: (1) the verification-to-target cost ratio $T_V(\gamma)/T_T$, (2) the draft-to-target cost ratio $T_D/T_T$, and (3) the expected accepted length $\Omega(\gamma,\alpha)$ for draft length $\gamma$ and acceptance rate $\alpha$.

In memory-bound LLM serving, the draft-to-target cost ratio and the expected acceptance rate become especially important. Lowering draft cost is closely related to reducing memory traffic during draft generation, whereas maintaining a high acceptance rate depends on preserving sufficient alignment between the draft and target models. In addition, the draft-to-target cost ratio varies across workloads because the relative contributions of weight traffic and KV-cache traffic change with batch size and sequence length \cite{sun2024triforce}. As a result, the most effective draft configuration is not fixed and can vary with the serving regime.

\subsection{Self-Speculative Decoding and the Need for Bit-Sharing Co-Design}

Because LLM serving is highly memory-intensive, the storage and bandwidth overhead of a separate draft model can significantly reduce the feasible serving range.
Self-speculative decoding avoids the need for a separately stored draft model by deriving draft predictions directly from the target model \cite{elhoushi2024layerskip, choi2026cassandra}.
As shown in Table ~\ref{tab:ar_sd_ssd}, this changes the role of the draft model from an explicitly stored network to an implicit low-cost representation derived from the target model itself.

Among self-speculative decoding methods, quantization-based bit-sharing approaches are particularly promising because they derive low-cost drafts directly from the target representation while preserving the original model structure \cite{liu2023online, yan2025decoding}.
Fig.~\ref{fig:b_s_efficiency} compares representative state-of-the-art approaches with BS-Spec across varying batch sizes and context lengths.
SPEQ~\cite{zhao2025quarter} constructs lightweight drafts using reduced floating-point information from the target representation. 
As shown in Fig.~\ref{fig:b_s_efficiency}(a), SPEQ achieves substantial speedup in short-context workloads, but its benefit diminishes as the context length increases. QuantSpec~\cite{tiwari2025quantspec} demonstrates that the preferred draft precision varies with the serving workload and that KV-cache quantization can effectively reduce draft cost. However, extending the same adaptive mechanism to weights requires separately stored quantized weight representations. We therefore evaluate both its fully self-speculative KV-only variant and its weight-non-self-speculative variant. As shown in Fig.~\ref{fig:b_s_efficiency}(b), the fully self-speculative QuantSpec performs well in long-context, small-batch workloads, but provides limited gains elsewhere. The weight-non-self-speculative QuantSpec achieves broader speedup, as shown in Fig.~\ref{fig:b_s_efficiency}(c), but its benefit diminishes in long-sequence, large-batch regimes. In contrast, BS-Spec achieves speedup across all evaluated workload regions, as shown in Fig.~\ref{fig:b_s_efficiency}(d), highlighting the need to jointly adapt weight and KV-cache precisions based on both draft cost and acceptance rate.

\subsection{Mixed-Precision Hardware for Bit-Shared Self Speculative Decoding}

Bit-shared speculative decoding constructs a low-precision draft model by reusing a subset of bits from the target model, enabling progressive refinement from low to full precision. As a result, a single hardware platform needs to support a wide range of bit precisions, from highly quantized draft execution to near full-precision target execution. In this setting, practical speedup depends not only on efficient memory access for selectively loading draft bits, but also on compute architectures that can execute low-precision drafts efficiently within the same hardware framework.

This motivates mixed-precision hardware that can adapt dynamically to changing bitwidths. Several prior works have explored this direction. SPEQ~\cite{zhao2025quarter}, for example, adopts a bit-shared self-speculative decoding approach related to BS-Spec, in which FP16 target weights and FP4 draft weights are co-designed. It stores low-precision weights separately and incrementally fetches additional bits for target execution. Its processing element (PE) includes multiple floating-point execution paths, allowing three FP units to operate in parallel during draft inference while using only a single FP unit for target computation. This design, however, still introduces inefficiencies: FP4 execution requires fetching 5 bits for decoding, and target-mode execution leaves part of the PE underutilized.

More broadly, prior mixed-precision accelerators also provide useful insights. BitFusion~\cite{sharma2018bit} proposes bit-flexible compute units, called BitBricks, that can be dynamically fused at runtime to support multiple bitwidths. This design supports efficient execution across a range of precisions, but is mainly oriented toward configurable fixed-precision execution rather than fine-grained bit sharing.

For more flexible precision support beyond power-of-two formats, bit-serial architectures offer an attractive alternative. Unlike conventional bit-parallel designs, bit-serial architectures process operands one bit-plane at a time, which naturally enables flexible precision scaling. Stripes \cite{judd2016stripes} uses this paradigm to support per-layer and per-bit precision control, improving both performance and energy efficiency. FIGLUT \cite{park2025figlut} further extends bit-serial computation by introducing LUT-based accumulation \cite{li2025lut, mo2025lut, im2024lutein}, replacing conventional MAC units with read-accumulate (RAC) units to support FP-INT and FP-BCQ operations efficiently while mitigating memory bank conflicts. In particular, it employs a flip-flop-based LUT (FFLUT) to effectively eliminate bank conflicts, demonstrating improved efficiency. LUT-SSM\cite{yoo202631} introduces a many-to-many LUT-ACC organization to improve LUT reuse for weight-only-quantized SSM execution and integrates element-wise operations into the same accelerator.

Despite these advances, existing architectures are not well matched to bit-shared speculative decoding. BS-Spec calls for flexible precision scaling across a broad range, together with efficient reuse of shared bits between draft and target execution. To support these requirements, we adopt a bit-serial load-and-compute paradigm that enables fine-grained bit-level control. At the same time, since naive bit-serial accumulation can introduce substantial overhead from serialized operations, BS-Spec integrates LUT-based accumulation to combine the flexibility of bit-serial computation with the efficiency of parallel accumulation.

\begin{table*}[]
\caption{Per-token acceptance rate for Llama-3.2-3B, Llama-3.1-8B, and Qwen3-14B model with various weight and KV cache bitwidth.}
{\scriptsize
\begin{tabularx}{1\textwidth}{c|ccccc|ccccc|ccccc}
\toprule[1.2pt]
Target       & \multicolumn{15}{c}{Llama-3.2-3B-Instruct BFP}
\\ 
\midrule
Draft weight & \multicolumn{5}{c|}{8}  & \multicolumn{5}{c|}{7}   & \multicolumn{5}{c}{6}
\\ 
\midrule
Draft KV     & \multicolumn{1}{c}{9} & \multicolumn{1}{c}{8} & \multicolumn{1}{c}{7} & \multicolumn{1}{c}{6} & \multicolumn{1}{c|}{5} & \multicolumn{1}{c}{9} & \multicolumn{1}{c}{8} & \multicolumn{1}{c}{7} & \multicolumn{1}{c}{6} & \multicolumn{1}{c|}{5} & \multicolumn{1}{c}{9} & \multicolumn{1}{c}{8} & \multicolumn{1}{c}{7} & \multicolumn{1}{c}{6} & \multicolumn{1}{c}{5} 
\\
\midrule
Humaneval    & 97.9\%                & 97.6\%                & 97.3\%                & 96.5\%                & 93.8\%                & 96.6\%                & 96.5\%                & 95.9\%                & 95.0\%                & 86.3\%                & 94.1\%                & 94.2\%                & 93.8\%                & 93.0\%                & 89.3\%                
\\
GSM8K        & 98.8\%                & 98.7\%                & 98.2\%                & 97.2\%                & 94.8\%                & 97.6\%                & 97.5\%                & 97.1\%                & 96.1\%                & 87.1\%                & 95.9\%                & 95.8\%                & 95.4\%                & 94.6\%                & 92.3\%               
\\
MBPP         & 98.7\%                & 98.6\%                & 98.3\%                & 97.3\%                & 95.5\%                & 97.5\%                & 97.2\%                & 97.1\%                & 96.3\%                & 88.2\%                & 94.9\%                & 94.9\%                & 94.8\%                & 94.0\%                & 92.0\%
\\
\midrule[1.2pt]

Target       & \multicolumn{15}{c}{Llama-3.1-8B-Instruct BFP} \\ \midrule
Draft weight & \multicolumn{5}{c|}{8}                                                                                                 & \multicolumn{5}{c|}{7}                                                                                                 & \multicolumn{5}{c}{6}                                                                                                 \\ \midrule
Draft KV     & \multicolumn{1}{c}{9} & \multicolumn{1}{c}{8} & \multicolumn{1}{c}{7} & \multicolumn{1}{c}{6} & \multicolumn{1}{c|}{5} & \multicolumn{1}{c}{9} & \multicolumn{1}{c}{8} & \multicolumn{1}{c}{7} & \multicolumn{1}{c}{6} & \multicolumn{1}{c|}{5} & \multicolumn{1}{c}{9} & \multicolumn{1}{c}{8} & \multicolumn{1}{c}{7} & \multicolumn{1}{c}{6} & \multicolumn{1}{c}{5} \\
\midrule
Humaneval & 98.1\% & 98.4\% & 98.0\% & 96.5\% & 92.6\% & 96.9\% & 96.7\% & 96.4\% & 95.2\% & 92.4\% & 94.3\% & 94.1\% & 93.8\% & 93.0\% & 90.1\% \\
GSM8K     & 98.6\% & 98.6\% & 98.1\% & 97.0\% & 94.5\% & 96.8\% & 96.7\% & 96.3\% & 95.3\% & 93.2\% & 93.2\% & 93.1\% & 92.8\% & 92.0\% & 89.5\% \\
MBPP      & 98.7\% & 98.6\% & 98.2\% & 97.5\% & 95.3\% & 97.4\% & 97.4\% & 97.0\% & 96.3\% & 94.5\% & 94.6\% & 94.4\% & 94.2\% & 93.5\% & 91.1\% \\
\midrule[1.2pt]
Target       & \multicolumn{15}{c}{Qwen3-14B BFP}   \\
\midrule

Draft weight & \multicolumn{5}{c|}{7}                                                                                                 & \multicolumn{5}{c|}{6}                                                                                                 & \multicolumn{5}{c}{5}                                                                                                 \\ \midrule
Draft KV     & 9      & 8      & 7      & 6      & 5      & 9      & 8      & 7      & 6      & 5      & 9      & 8      & 7      & 6      & 5      \\
\midrule
Humaneval    & 97.7\% & 97.6\% & 97.3\% & 95.5\% & 91.8\% & 95.5\% & 95.6\% & 95.1\% & 94.3\% & 91.1\% & 91.2\% & 91.1\% & 91.2\% & 90.5\% & 88.0\% \\
GSM8K        & 98.3\% & 98.2\% & 97.9\% & 97.0\% & 94.3\% & 96.7\% & 96.8\% & 96.6\% & 95.8\% & 93.4\% & 93.7\% & 93.7\% & 93.6\% & 93.3\% & 91.4\% \\
MBPP         & 97.8\% & 97.6\% & 97.4\% & 96.0\% & 92.9\% & 95.9\% & 96.0\% & 95.8\% & 94.8\% & 91.8\% & 92.0\% & 92.0\% & 91.8\% & 91.5\% & 89.1\% \\
\bottomrule[1.2pt]
\end{tabularx}
}
\label{tab:acceptance_rate}
\end{table*}

\section{BS-Spec Algorithm}

\subsection{BFP Target Representation}

\textcolor{black}{
BS-Spec stores the weight and KV cache of the target model in BFP and loads only a subset of their bits during the draft phase. Fig.~\ref{fig:bfp_make} illustrates the BFP conversion process, in which each group is represented by aligned significands and a shared exponent \cite{jang2024figna}. BFP representation enables two important optimizations in self-speculative decoding:  (1) hardware-friendly draft construction and (2) high acceptance rates.
Speculative decoding requires fast draft generation, as any additional computation in the draft phase reduces the achievable speedup. Therefore, hardware-friendly drafting should construct the draft model directly from the target model without additional processing overhead. At the same time, because the speedup strongly depends on the acceptance rate, the partially loaded draft model is required to remain numerically close to the full target model.
}

\textcolor{black}{
Fig. ~\ref{fig:numerical_example} presents a numerical example of 7-bit partial loading under different schemes. Partial loading directly from BF16 substantially distorts the original value, particularly when only mantissa bits are retained. Although exponent-based loading preserves the numerical scale better than mantissa-based loading, the proposed BFP-based scheme retains both the aligned significand and the shared exponent, resulting in the closest approximation to the original value. The example uses a group size of three for clarity, which makes the exponent overhead appear relatively large. In practice, a larger group size, such as 64, amortizes this overhead across many values. 
}

\textcolor{black}{
This partial-loading approach can also be implemented in a hardware-friendly manner. During the draft phase, only a subset of the target model bits is loaded, reducing memory traffic and alleviating the memory-bound bottleneck. Moreover, BFP naturally supports multiple draft precisions through partial-bit loading, making it particularly well suited to self-speculative decoding. Consequently, storing the target model in BFP and using its partial-bit representations for drafting provides an efficient way to reduce draft cost while maintaining a high acceptance rate.}



For each group, BS-Spec selects a shared exponent based on the maximum exponent in the group and aligns all values to that exponent through right shifting. Weights are quantized offline and stored in BFP9 format. KV-cache values, in contrast, are quantized online during inference: keys are grouped channel-wise, whereas values are grouped token-wise. Because channel-wise grouping for keys cannot always be finalized at every decoding step, BS-Spec temporarily buffers key values until a full group is formed and then converts them to BFP. BS-Spec applies BFP9 only to weights and the KV cache, while keeping activations in the original BF16 format.

\subsection{Self-Speculative Decoding with BFP}

\textcolor{black}{Instead of storing a separate draft model, BS-Spec derives the draft path directly from the BFP target representation by reducing the mantissa precision of weights, the KV cache, or both, while keeping the target path unchanged for verification.
Fig.~\ref{fig:overview} illustrates the detailed inference flow of BS-Spec.  During drafting, BS-Spec selects a reduced-precision pair $(p_w, p_{kv})$, where $p_w$ and $p_{kv}$ denote the mantissa widths used for weights and the KV cache, respectively. Based on this choice, the draft path reads only the corresponding high-order mantissa bits from the stored target representation and generates $\gamma$ candidate tokens. At the same time, temporary KV entries are created for the drafted sequence. The target path then verifies these candidates using the full BFP representation, and only accepted tokens are committed to the persistent model state.
}

Because weight and KV-cache precision can be selected independently through $(p_w, p_{kv})$, BS-Spec is not tied to a single fixed draft model. Instead, it instantiates different draft paths directly from the same stored representation according to the selected precision pair.

\subsection{Trade-Offs in Draft Model Precision}

\begin{figure}[t]
    \centering
    \includegraphics[width=0.49\textwidth]{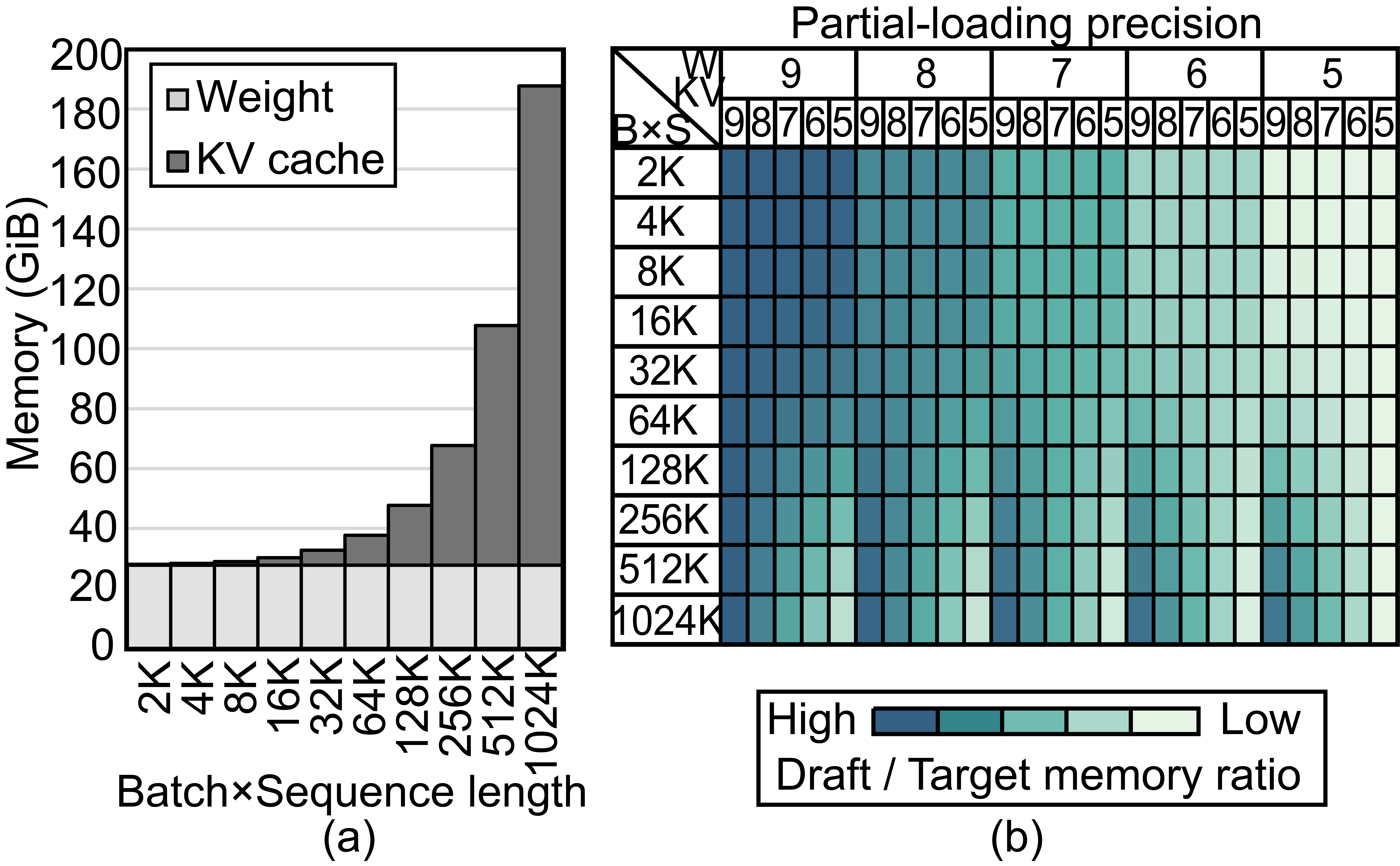}
    \caption{\textcolor{black}{For Qwen3-14B across varying batch size $\times$ sequence length: (a) the memory footprints of the weights and KV cache, and (b) the impact of weight and KV cache precision reduction on the Draft / Target memory ratio.}}
    \label{fig:weight_kv_ratio}
\end{figure}


\textcolor{black}{
To efficiently exploit partially loaded draft models, BS-Spec needs to select appropriate weight and KV-cache precisions for each runtime inference configuration. The optimal precision pair is determined by two factors: the draft-model cost and the acceptance rate.
}

\textbf{Draft Model Cost.}
In speculative decoding, weight and KV-cache compression are often considered independently. However, weights and the KV cache exhibit fundamentally different scaling behaviors in practical LLM inference. The weight footprint remains fixed once the model architecture is determined, whereas the KV-cache footprint grows linearly with batch size and sequence length. Consequently, their relative contributions to memory traffic vary with the serving configuration, leading to different optimal weight and KV-cache precisions across batch sizes and sequence lengths.
Fig.~\ref{fig:weight_kv_ratio} illustrates this trend for Qwen3-14B under varying batch size $\times$ sequence length.
Fig.~\ref{fig:weight_kv_ratio}(a) compares the memory footprints of the weights and KV cache. At small batch sizes and short sequence lengths, the weight footprint dominates, whereas the contribution of the KV cache increases as either dimension grows.
Fig.~\ref{fig:weight_kv_ratio}(b) evaluates all combinations of 5--9-bit weight and KV-cache precisions for the proposed BFP target model and its partially loaded draft model. Each color represents the draft-to-target memory-loading ratio. 
Darker colors indicate that partial loading provides only a limited reduction in draft latency, whereas lighter colors represent lower-cost draft models with substantially reduced memory traffic. These results show that weight and KV-cache precisions must be selected jointly according to the current batch size and sequence length.

\textbf{Acceptance rate.}
Reducing the draft-to-target ratio lowers the draft cost but generally increases the numerical difference between the draft and target models, thereby reducing the acceptance rate. Conversely, a larger draft-to-target ratio typically improves the acceptance rate at the expense of a more costly draft phase. Thus, speculative-decoding performance is governed by a trade-off between draft latency and acceptance rate, and the precision combinations that maximize speedup form a Pareto frontier over these two factors.
Table ~\ref{tab:acceptance_rate} reports the per-token acceptance rates for representative range of 5--9-bit weight and KV-cache precisions. The results are evaluated on HumanEval, GSM8K, and MBPP using Llama-3.2-3B, Llama-3.1-8B, and Qwen3-14B. Table~\ref{tab:speq_qs_acceptance} compares the average per-token acceptance rates of SPEQ\cite{zhao2025quarter} and three QuantSpec\cite{tiwari2025quantspec} draft configurations. We evaluate $\gamma$ from 5 to 8 and $\gamma=7$ is selected, which provides the largest expected accepted generation length on average. By combining the estimated draft cost with the measured acceptance rate for each precision pair, BS-Spec identifies the configuration that maximizes the expected speculative-decoding speedup.

%

\begin{table}[t]
\centering
\caption{Average per-token acceptance rates across HumanEval, GSM8K, and MBPP for SPEQ~\cite{zhao2025quarter} and three QuantSpec~\cite{tiwari2025quantspec} draft configurations on Llama-3.2-3B, Llama-3.1-8B, and Qwen3-14B.}
\begin{tabularx}{0.47\textwidth}{c|c|ccc}
\toprule[1.2pt]
& SPEQ   & \multicolumn{3}{c}{QuantSpec} \\
             & W4KV16 & W4KV8    & W4KV4    & W16KV4  \\
             \midrule
Llama-3.2-3B & 87.4\% & 95.6\%   & 87.2\%   & 90.3\%  \\
Llama-3.1-8B & 86.1\% & 96.2\%   & 88.2\%   & 91.3\%  \\
Qwen3-14B    & 89.8\% & 96.0\%   & 93.9\%   & 93.9\% 
\\
\bottomrule[1.2pt]

\end{tabularx}
\label{tab:speq_qs_acceptance}
\end{table}

\subsection{Runtime Draft Precision Selection for the Best Speedup}

\begin{figure}[t]
    \centering
    \includegraphics[width=0.48\textwidth]{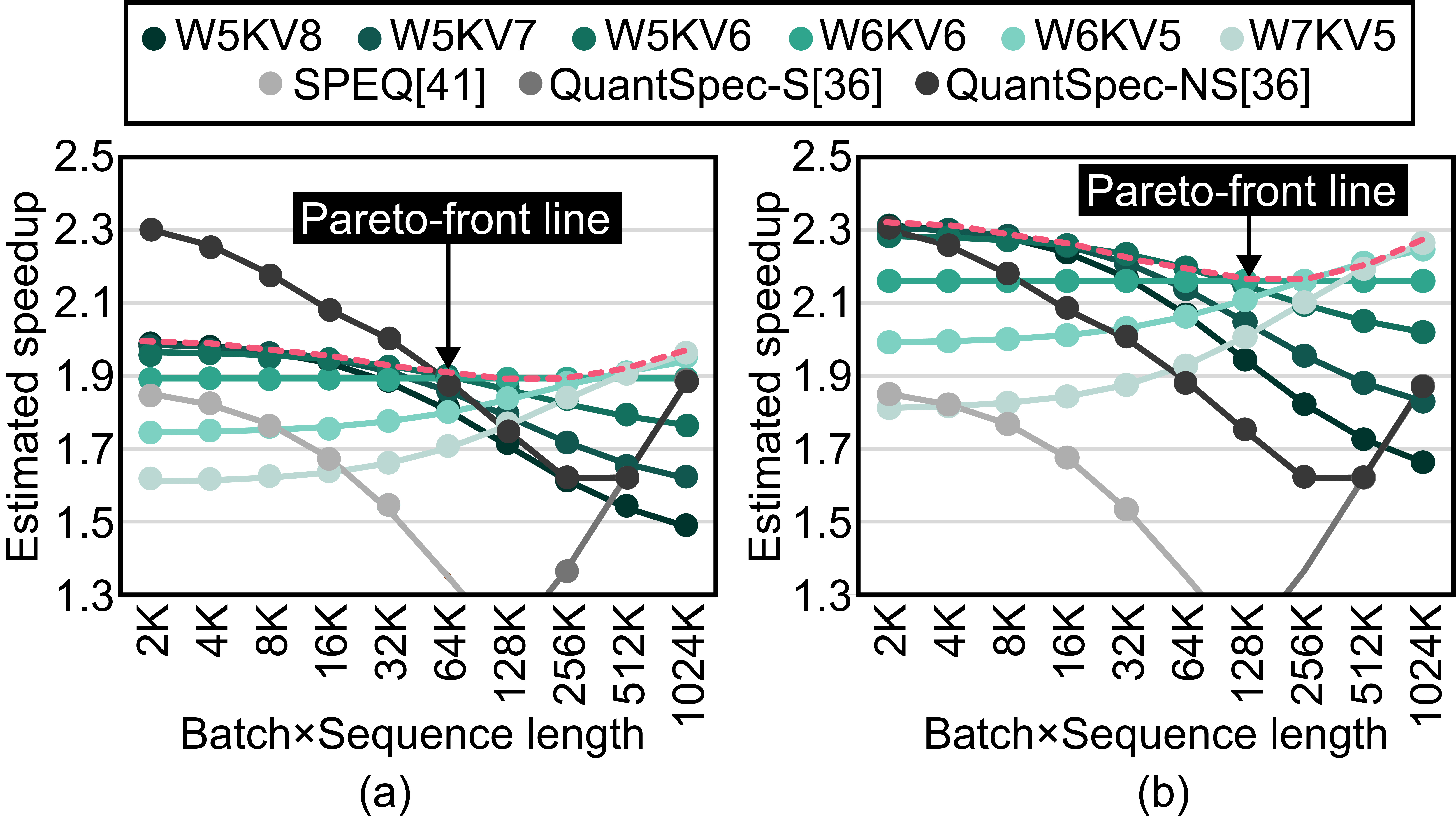}
    \caption{Estimated speedup of Qwen3-14B over the non-speculative 16-bit baseline. Speculative decoding is applied in both cases, using (a) the original BFP16 target model and (b) a BFP9 target model.}
    \label{fig:pareto_front}
\end{figure}

\begin{table}[t]
\centering
\caption{Pareto-optimal configurations stored in the precision lookup table, selected to maximize the average speedup for each model across varying batch size$\times$ sequence length.}
\begin{tabular}{c|ccc}
\toprule[1.2pt]
       & Llama-3.2-3B & Llama-3.1-8B & Qwen3-14B \\
       \midrule
2K    & W6KV8        & W6KV9        & W5KV8     \\
4K    & W6KV8        & W7KV9        & W5KV8     \\
8K    & W6KV7        & W7KV8        & W5KV7     \\
16K   & W6KV6        & W7KV7        & W5KV7     \\
32K   & W6KV6        & W7KV7        & W5KV6     \\
64K   & W6KV6        & W7KV6        & W5KV6     \\
128K  & W7KV5        & W7KV6        & W6KV6     \\
256K  & W8KV5        & W8KV5        & W6KV6     \\
512K  & W8KV5        & W8KV5        & W6KV5     \\
1024K & W8KV5        & W8KV5        & W7KV5     \\
\bottomrule[1.2pt]
\end{tabular}
\label{tab:pareto_front_options}
\end{table}


Using two factors, we estimate the speculative decoding speedup and identify the Pareto-optimal precision configurations.
Equations~\ref{eq:weight_mem} and~\ref{eq:kv_mem} describe the memory footprint of weights and KV cache, respectively.

{\small
\begin{equation}
    MEM_W = Bytes \cdot N_{Param}
    \label{eq:weight_mem}
\end{equation}
\begin{equation}
    MEM_{KV} = 2 \cdot Bytes \cdot N_{Layer} \cdot (Batch \cdot N_{KV} \cdot T \cdot D_h)
    \label{eq:kv_mem}
\end{equation}
}

\textcolor{black}{
Under the memory-bound inference regime targeted by speculative decoding, the draft-to-target memory loading ratio approximates the draft-to-target latency ratio, $T_D/T_T$. As a result, in the BFP-based bit-sharing speculative decoding scenario, the modified speedup formula is written as follows:
}

{\small
\begin{equation}
    \frac{T_{Avg}^{SD}}{T_T}=\frac{1}{\Omega (\gamma, \alpha)} \left( \gamma \frac{ p_{w} R_{w}(BS)+ p_{kv}R_{kv}(BS)}{p_{full}} + \frac{T_V(\gamma)}{T_T} \right) 
\end{equation}
\label{eq:BIS_Spec_speedup}
}

Here, $R_w$ and $R_{kv}$ denote the weight- and KV cache-loading ratios, respectively, as functions of batch size $\times$ sequence length, denoted by $BS$.

\textcolor{black}{
Using this formulation, the estimated speedup can be evaluated for every weight and KV-cache precision pair. Fig.~\ref{fig:pareto_front} presents the resulting offline calibration for Qwen3-14B. Each precision configuration produces an estimated speedup curve over varying batch size $\times$ sequence length, and the configuration achieving the highest estimated speedup at each operating point is selected. The dotted line represents the upper envelope formed by these optimal configurations.
Table~\ref{tab:pareto_front_options} summarizes the selected precision configurations across the evaluated benchmarks.
Fig.~\ref{fig:pareto_front} further compares two target-model precision settings. 
Fig.~\ref{fig:pareto_front}(a) reports the estimated speedup over a 16-bit autoregressive baseline when BS-Spec uses a BFP16 target model. 
Fig.~\ref{fig:pareto_front}(b) considers a BFP9 target model, thereby accelerating not only the draft phase but also target verification. BFP9 retains a 9-bit aligned significand consisting of the sign bit, implicit leading one, and seven BF16 mantissa bits, without an additional guard-bit margin.
Because BFP9 reduces the target-model latency, the normalized draft cost $T_D/T_T$ becomes larger, which can reduce the incremental speedup obtained purely from speculative decoding relative to the target model. Nevertheless, accelerating the target model itself more than compensates for this reduction, yielding a higher estimated overall speedup relative to the 16-bit baseline.
Moreover, Table~\ref{tab:accuracy} shows that the BFP9 target preserves the accuracy of the BF16 target across the evaluated workloads. Therefore, BS-Spec uses the BFP9 target with its partially loaded draft representations.
}

Through this offline estimation process, BS-Spec identifies the weight and KV-cache precision pair that maximizes predicted speedup for each batch-size and context-length configuration and stores it in the PLT. At runtime, the selected precision pair is retrieved from the PLT and directly mapped onto the BS-Spec hardware for execution.


\begin{table}[t]
\caption{GSM8K accuracy of Llama-3.2-3B, Llama-3.1-8B, and Qwen3-14B under the original BF16 baseline, a SPEQ\cite{zhao2025quarter} full precision FP16 model, QuantSpec\cite{tiwari2025quantspec} INT8 KV model, and the BS-Spec target model.}
\begin{tabularx}{0.48\textwidth}{cc|ccc}
\toprule[1.2pt]
Model & Option & 3B &8B &14B  \\
\midrule
Original      & BF16 W, BF16 KV & 77.7\% &84.5\% &91.4\%        \\
SPEQ      & FP16 W, FP16 KV & 77.7\% &84.5\% &91.4\%        \\
QuantSpec & BF16 W, INT8 KV & 76.8\%  &84.8\% &91.1\%       \\
BS-Spec     & BFP9 W, BFP9 KV & 77.8\% &85.5\% &91.0\%        \\
\bottomrule[1.2pt]
\end{tabularx}

\label{tab:accuracy}
\end{table}



 \begin{figure}[t]
     \centering
     \includegraphics[width=0.5\textwidth]{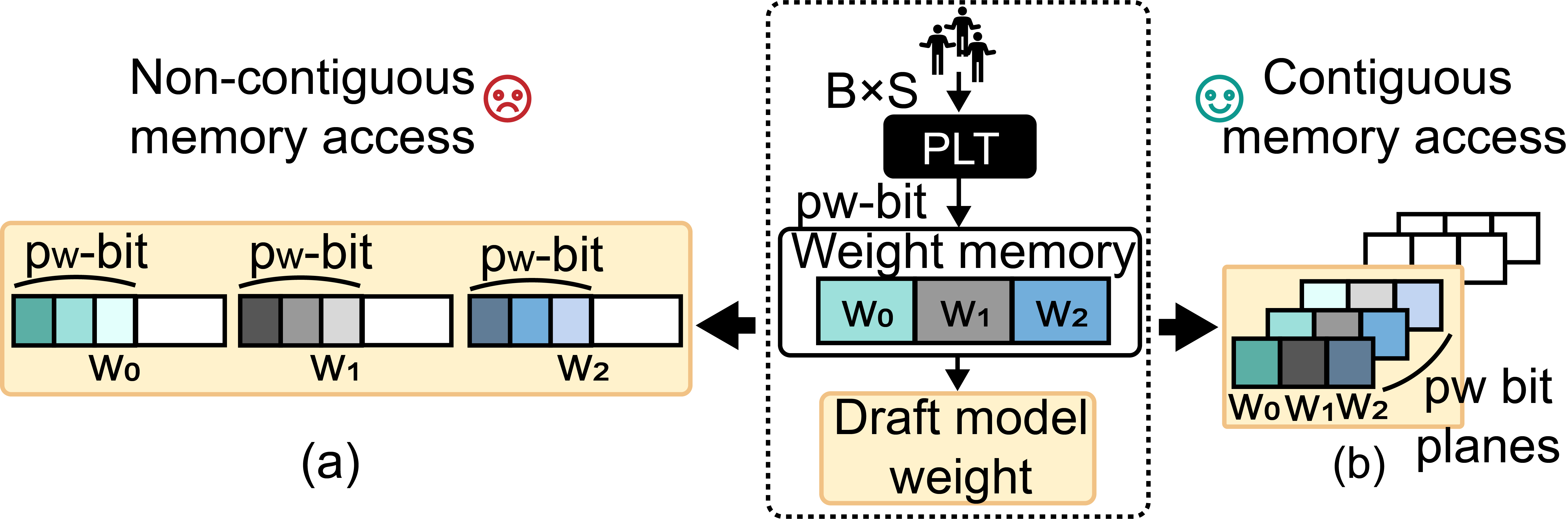}
     \caption{Comparison of data layout and memory-load for loading $p_w$-bit weights in BS-Spec using (a) bit-parallel and (b) bit-serial organizations; the same organization is applied to $p_{kv}$-bit KV-cache loading.}
     \label{fig:bit_serial_bit_parallel}
 \end{figure}

 
\section{Hardware Support for Bit-Shared BFP Execution}


\subsection{Variable-Precision Bit-Plane Execution}

BS-Spec reduces draft-generation cost by selectively loading only the MSB planes of BFP-formatted weights and KV-cache entries. However, the benefits of partial loading can be fully realized only when the memory format is adequate and efficiently uses the bandwidth. The accelerator must also support both low-precision draft execution and higher-precision target verification using the same datapath.

To support variable-precision datapath, BS-Spec adopts bit-plane major architecture. Fig.~\ref{fig:bit_serial_bit_parallel} compares (a) conventional storage and (b) the bit-plane-major memory organization used by BS-Spec. Instead of storing all bits of one value consecutively, bits of the same significance from multiple values are placed contiguously \cite{park2025anybcq}. 
For draft generation, the controller issues $p_w$-bit weight accesses and $p_{kv}$-bit KV-cache accesses, selected by the PLT, from MSB planes to lower orders. Only these selected bit planes are transferred from memory and processed by the compute engine. During target verification, the same datapath executes the full target precision by extending the bit-serial iteration over the remaining bit planes. Thus, draft and target computation share a common representation, memory organization, and arithmetic datapath; changing precision requires only modifying the number of loaded and executed bit planes.

This design ensures that precision reduction benefits both the memory and compute sides. Reducing $p_w$ or $p_{kv}$ decreases the number of bit planes fetched from off-chip memory and proportionally reduces the number of bit-serial accumulation iterations. Therefore, the runtime precision selected by the algorithm is directly translated into lower data movement and execution cost, rather than being converted into an internally fixed-precision format.

\subsection{LUT-driven dataflow}
BS-Spec adopts LUT-based accumulation to efficiently compute products between BF16 activations and bit-serial BFP weights or KV-cache values. The basic computation follows LUT-GEMM \cite{park2024lut}: groups of activations are used to precompute partial sums for all possible binary weight patterns, and multiplication is replaced with indexed LUT access followed by accumulation. 

Each LUT is connected to multiple read-accumulate lanes. In every cycle, streamed weight bits form the LUT indices, and the selected entries are added to local partial sums with the appropriate bit-position shift. Since bit planes are processed serially, the number of execution cycles scales approximately with the selected precision. And LUT construction is pipelined with execution so that the LUT for the next activation group can be generated while the current group is being processed, hiding most of the generation latency. 
As derived as optimal option in \cite{park2025figlut}, BS-Spec uses an index width of $\mu=4$ and RAC unit fanout of 32.


 \begin{figure}[t]
     \centering
     \includegraphics[width=0.48\textwidth]{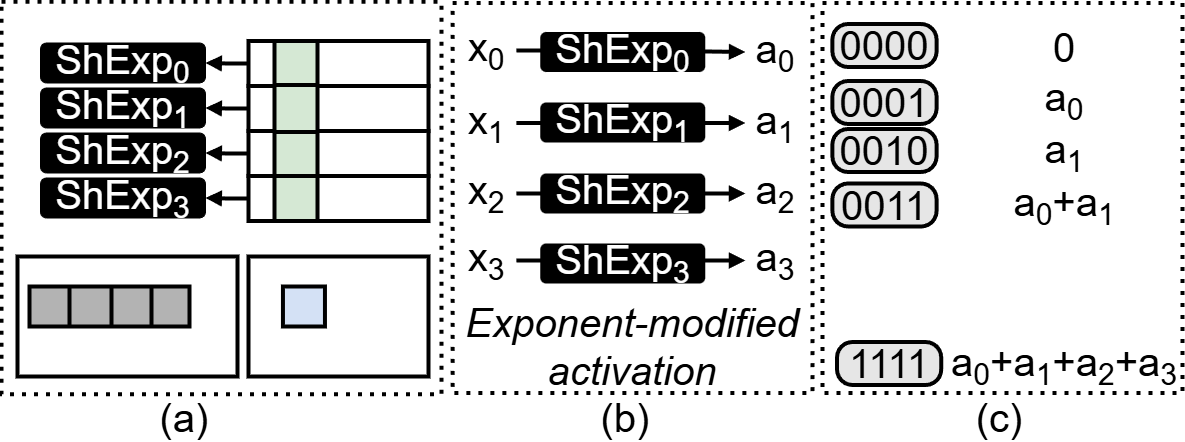}
     \caption{(a) Desired computation, (b) exponent-modified activation, and (c) LUT index and corresponding activation sum.}
     \label{fig:lut_gen}
 \end{figure}

  \begin{figure}[t]
     \centering
     \includegraphics[width=0.5 \textwidth]{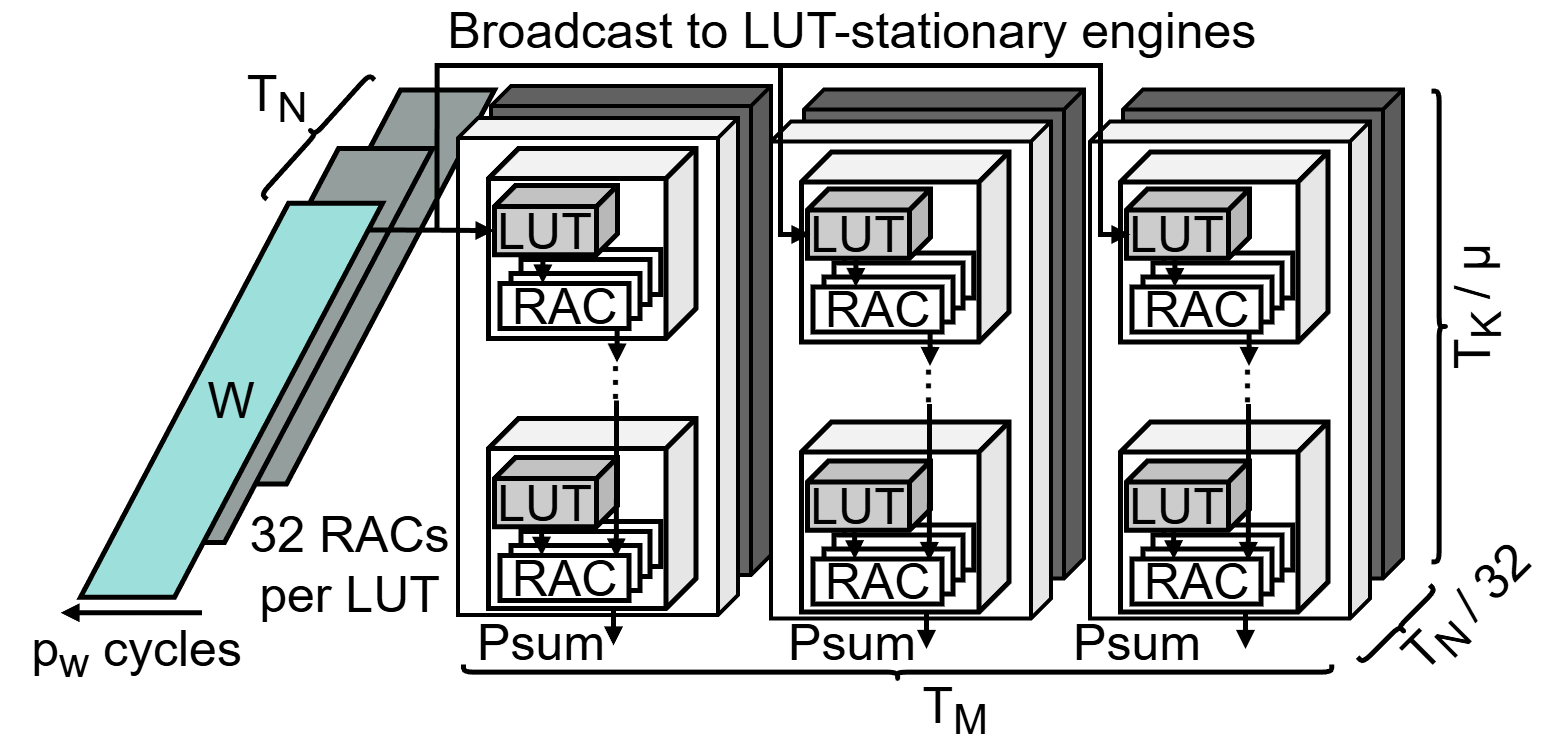}
     \caption{LUT-stationary dataflow of BS-Spec, where weight bit planes flow through the datapath and are broadcast across the engines. The KV cache follows the same dataflow.}
     \label{fig:lut_stationary}
 \end{figure}
 
 \begin{figure}[t]
     \centering
     \includegraphics[width=0.45\textwidth]{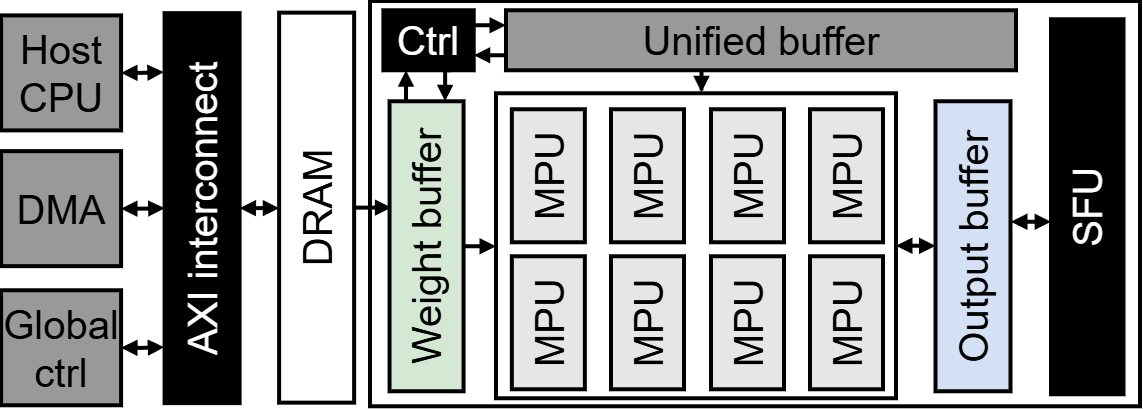}
     \caption{Integrated system architecture of BS-Spec.}
     \label{fig:system_architecture}
 \end{figure}

  \begin{figure*}[t]
     \centering
     \includegraphics[width=0.99\textwidth]{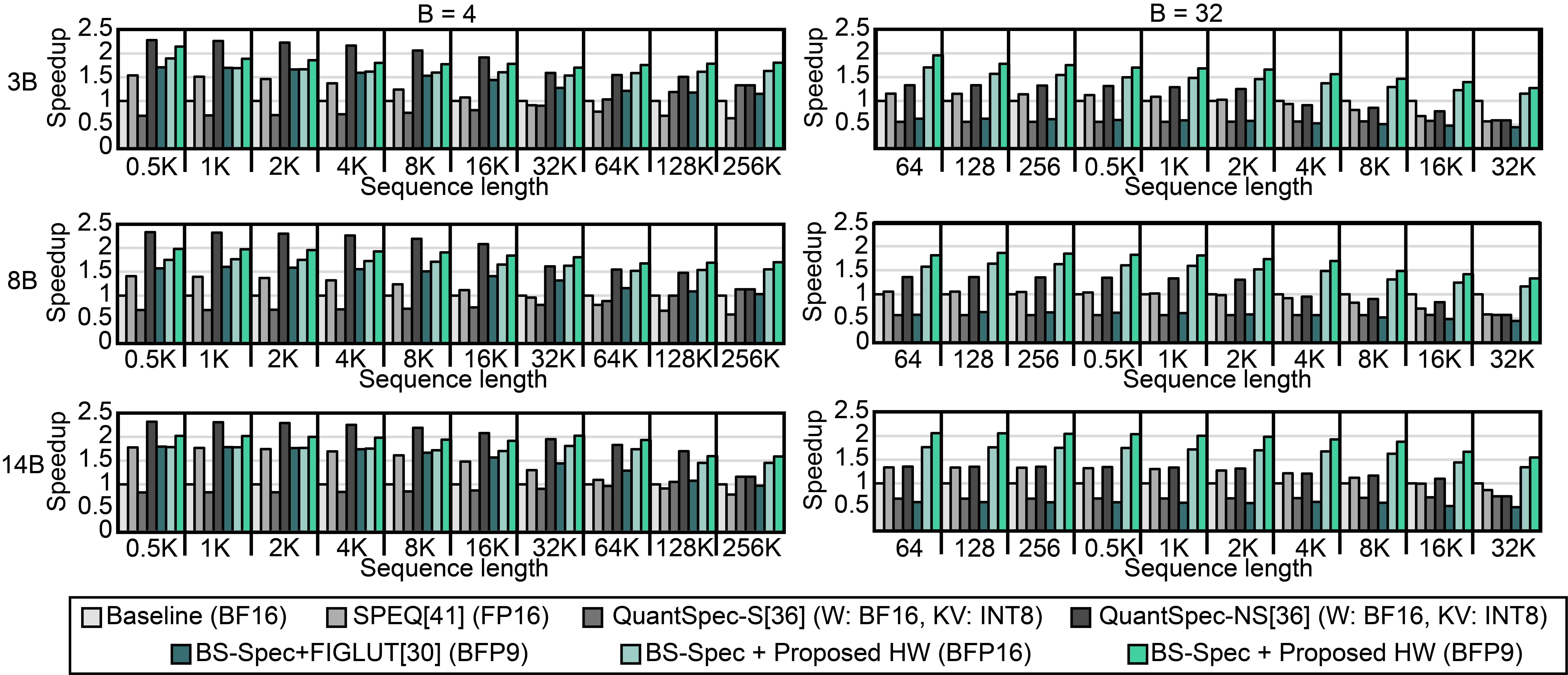}
     \caption{Comparison of three methods in terms of normalized speedup for batch (B) 4 and 32, across three models: Llama-3.2-3B, Llama-3.1-8B, and Qwen3-14B. Baseline (value = 1) corresponds to target-only autoregressive decoding on a GPU-style systolic array.}
     \label{fig:normalized_speedup}
 \end{figure*}

\subsection{BFP-Aware LUT Computation}

The proposed LUT generator is modified to directly incorporate the BFP shared exponent. For each weight group, the activations are first scaled according to the corresponding shared exponent and shifted to a common aligned representation, as shown in Fig.~\ref{fig:lut_gen}. The aligned activation mantissas are then used to construct LUT entries for all possible weight-bit combinations. As a result, exponent alignment is folded into LUT precomputation, and the original floating-point dot product is transformed into a LUT lookup followed by integer accumulation.

Because BS-Spec stores BFP significands in sign-magnitude form, the sign bit is handled separately from the remaining magnitude bit planes. A lightweight sign decoder interprets the sign plane and conditionally selects the positive or negative contribution associated with the corresponding LUT index. The remaining magnitude planes are processed from the most significant to the least significant selected bit, with the lookup result shifted according to the bit position before integer accumulation.

\subsection{Decoding-Oriented Dataflow}
To efficiently support the autoregressive decoding phase, BS-Spec adopts a LUT-stationary datapath. For a tiled multiplication between $A[T_M \times T_K]$ and $W[T_K \times T_N]$, $T_M$ is typically small. For linear layers, $T_M$ is determined primarily by the serving batch size. For grouped-query attention, the corresponding matrix operations may expose only the number of query heads sharing each KV head as the effective $T_M$. Consequently, the reuse opportunity obtained by retaining weights can be limited in the decoding workloads targeted by speculative decoding method.

Fig.~\ref{fig:lut_stationary} shows the hardware mapping of BS-Spec. LUT entries generated from an activation group remain local to each MPU, while the corresponding weight bit planes are streamed through the read-accumulate lanes. The streamed weight bits serve as LUT indices, and the resulting partial sums are accumulated locally. Weight bit planes are broadcast across the MPUs, whereas LUT entries and partial sums remain private to each MPU. Each MPU independently retains and updates its partial sums, allowing multiple output tiles to be computed concurrently and operates in SIMD-style organization.



\subsection{System Integration}

Fig.~\ref{fig:system_architecture} shows the overall architecture of the BS-Spec accelerator. The system consists of a host processor, a global controller, a memory subsystem, and a BS-Spec compute core. The host processor issues high-level execution commands, while the global controller handles runtime scheduling, configuration, and coordination of data movement. These components are connected through a system interconnect (e.g., AXI), which also provides access to off-chip DRAM. Data transfer between DRAM and the compute core is managed by a DMA engine, which fetches weights, activations, and KV-cache data and streams them into the on-chip buffers. This organization decouples memory access from computation and allows data transfer to overlap with execution.

The BS-Spec core contains a unified buffer, a weight buffer, multiple MPUs, an output buffer, and a special function unit (SFU) for post-processing operations. The unified buffer stores activations and intermediate data, while the weight buffer supplies weights to the compute units. Computation is carried out by the MPUs, which implement SIMD-style tiled matrix multiplication.

\section{Experimental Results}

 \begin{figure*}[t]
     \centering
     \includegraphics[width=0.99\textwidth]{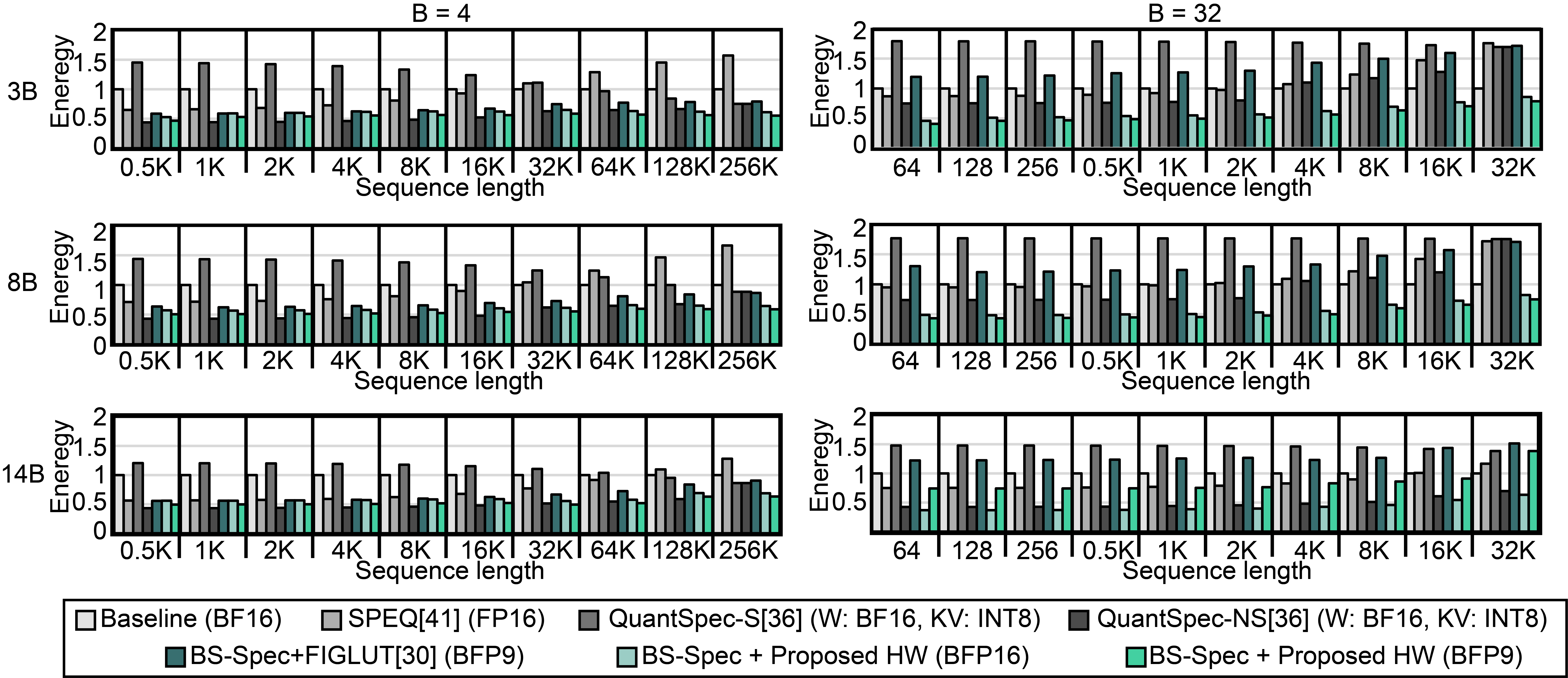}
     \caption{Comparison of three methods in terms of normalized energy for batch (B) 4 and 32, across three models: Llama-3.2-3B, Llama-3.1-8B, and Qwen3-14B. Baseline (value = 1) corresponds to target-only autoregressive decoding on a GPU-style systolic array.}
     \label{fig:normalized_energy}
 \end{figure*}

\begin{table}[t]
\centering
\caption{Comparison of computation unit area in the evaluated environment.}
\begin{tabularx}{0.4\textwidth}{c|ccc}
\toprule[1.2pt]
& SPEQ\cite{zhao2025quarter}  & QuantSpec\cite{tiwari2025quantspec} & BS-Spec                            
\\
\midrule
Area ($\mu m^2$) & \multicolumn{1}{c}{$5.56\times10^7$} & \multicolumn{1}{c}{$3.62\times10^7$} & \multicolumn{1}{c}{$2.51\times10^7$}
\\
\bottomrule[1.2pt]
\end{tabularx}
\label{tab:area}
\end{table}

\subsection{Hardware Evaluation Setup}

We implement the proposed BS-Spec accelerator in a 28-nm CMOS technology and target a clock period of 3\,ns (333\,MHz). Arithmetic units are synthesized using Synopsys DesignWare libraries, while on-chip SRAM parameters are obtained from CACTI \cite{balasubramonian2017cacti}. Off-chip HBM2 latency and energy are modeled using Ramulator 2 \cite{luo2023ramulator} with workload-generated memory traces. 
We model an HBM2 system based on~\cite{jun2017hbm} and assume access power of 10.0 W. All evaluated architectures use the same memory hierarchy and bandwidth constraints.

We compare BS-Spec with QuantSpec \cite{tiwari2025quantspec} and SPEQ \cite{zhao2025quarter} in terms of effective latency and energy consumption. Since QuantSpec is originally implemented as a GPU kernel without a dedicated accelerator, we map it onto a synthesized multi-precision systolic-array architecture supporting FP16--FP16, FP16--INT8, and dual-lane FP16--INT4 execution. The integer datapaths reuse the mantissa multipliers in the FP16 units to reduce hardware overhead.

Both QuantSpec and SPEQ are implemented with systolic-array architectures comprising four engines, each containing a $64\times64$ array. The number of instantiated arrays is scaled according to the effective bit-level parallelism to match the normalized output throughput. 
BS-Spec adopts a SIMD architecture consisting of 32 MPUs along the $T_M$ dimension, where
each MPU contains $T_N/32=2$ LUT--RAC subarrays. Each subarray
contains $T_K/\mu=16$ LUTs, with each LUT shared by 32 BS-RAC lanes.
Weights are broadcast across all MPUs, while LUT entries and partial sums remain local.
Table~\ref{tab:area} compares the resulting compute-unit areas, confirming that the evaluation does not favor BS-Spec through disproportionate compute provisioning.


\begin{figure}[t]
    \centering
    \includegraphics[width=0.49\textwidth]{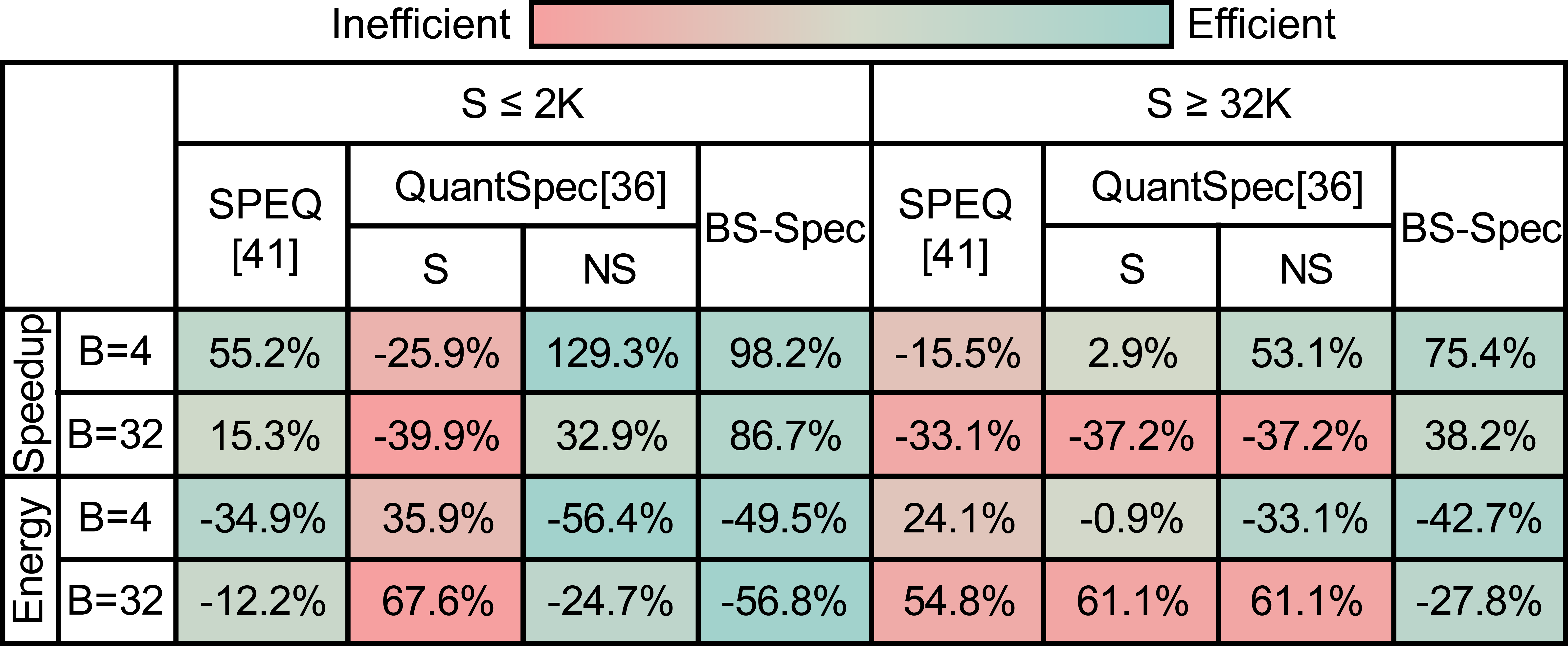}
    \caption{Average speedup and normalized energy consumption of the evaluated methods across short- and long-context regimes at batch sizes 4 and 32.}
    \label{fig:comparison_short_long}
\end{figure}
 
\subsection{Speedup and Energy Evaluation}



We report normalized effective throughput and energy consumption using the speculative-decoding formulation in Eq.~\ref{eq:sd_speedup}. The baseline is conventional autoregressive decoding of the BF16 target model on the systolic-array architecture described above. All results are normalized to the throughput and per-token energy of this baseline.

The normalized effective throughput is modeled as follows:
\begin{equation}
    \frac{Th_{\mathrm{SD}}}{Th_B}
    =
    \frac{
        \Omega(\gamma,\alpha) /
        \left(\gamma T_D + T_T\right)
    }{
        (\gamma+1) /
        \left((\gamma+1)T_B\right)
    }
    =
    \frac{
        \Omega(\gamma,\alpha) T_B
    }{
        \gamma T_D + T_T
    },
\end{equation}
where $T_D$, $T_T$, and $T_B$ denote the draft, target-verification, and baseline per-token latencies, respectively.

Similarly, the normalized energy consumption per generated token is modeled as:
\begin{equation}
    \frac{E_{\mathrm{SD}}}{E_B}
    =
    \frac{
        \left(\gamma E_D + E_T\right) /
        \Omega(\gamma,\alpha)
    }{
        (\gamma+1)E_B / (\gamma+1)
    }
    =
    \frac{
        \gamma E_D + E_T
    }{
        \Omega(\gamma,\alpha) E_B
    },
\end{equation}
where $E_D$, $E_T$, and $E_B$ denote the corresponding energy costs.
All latency and energy terms are obtained from cycle-level modeling that includes computation, SRAM and HBM accesses, and pipeline overlap under identical memory-bandwidth constraints.

We compare SPEQ~\cite{zhao2025quarter}, QuantSpec~\cite{tiwari2025quantspec}, and BS-Spec in terms of normalized throughput and energy. For QuantSpec, QuantSpec-S is fully self-speculative, storing only the full-precision target model and using a W16KV4 draft, whereas QuantSpec-NS additionally stores INT4 weights and adaptively selects among W4KV8, W4KV4, and W16KV4. For BS-Spec, we evaluate a FIGLUT-style mapping~\cite{park2025figlut}, the proposed hardware with a BFP16 target, and the complete design with a BFP9 target.
Fig.~\ref{fig:normalized_speedup} compares normalized throughput across Llama-3.2-3B, Llama-3.1-8B, and Qwen3-14B at batch sizes 4 and 32. At batch size 4, the complete BS-Spec design achieves speedups of 1.83$\times$, 1.84$\times$, and 1.90$\times$ over the autoregressive baseline, respectively; at batch size 32, the corresponding speedups are 1.62$\times$, 1.68$\times$, and 1.92$\times$.
Fig.~\ref{fig:normalized_energy} reports normalized energy under the same settings. At batch size 4, BS-Spec consumes 0.55$\times$, 0.54$\times$, and 0.53$\times$ the baseline energy, respectively; at batch size 32, the corresponding values are 0.55$\times$, 0.52$\times$, and 0.43$\times$.

QuantSpec-NS achieves competitive throughput and energy efficiency in some small-batch, short-context workloads by using a separately stored INT4 weight-draft model with a high acceptance rate. However, this benefit requires additional weight storage. In contrast, BS-Spec maintains high efficiency over a broader range of workloads while remaining fully self-speculative.
The comparison among the three BS-Spec configurations further illustrates the contributions of the proposed hardware and target compression. Compared with the FIGLUT-style LUT-based bit-serial architecture, the proposed hardware provides higher throughput and lower energy consumption when executing the BS-Spec algorithm. Compressing the target representation from BFP16 to BFP9 provides an additional improvement by reducing both draft and target execution costs.

To further summarize the workload-dependent trends, Fig.~\ref{fig:comparison_short_long} reports the average speedup and normalized energy consumption in short-context ($S \leq 2\mathrm{K}$) and long-context ($S \geq 32\mathrm{K}$) regimes. At batch size 4, SPEQ is more effective in the short-context regime, whereas QuantSpec becomes relatively more competitive at long contexts as KV-cache traffic accounts for a larger fraction of execution cost. At batch size 32, the higher arithmetic intensity changes the relative behavior of the evaluated methods by partially reducing the impact of memory traffic. Because QuantSpec targets execution on a conventional systolic-array organization, these results further suggest that its fine-grained adaptive precision policy can benefit from dedicated hardware support tailored to variable-precision speculative decoding.
Across both context regimes, BS-Spec achieves the highest overall efficiency by jointly adapting weight and KV-cache precisions to the workload while retaining a unified bit-sharing method and a common hardware datapath for all draft and target configurations.


\subsection{BS-RAC Power and Area Breakdown}

\begin{figure}[t]
     \centering
     \includegraphics[width=0.49\textwidth]{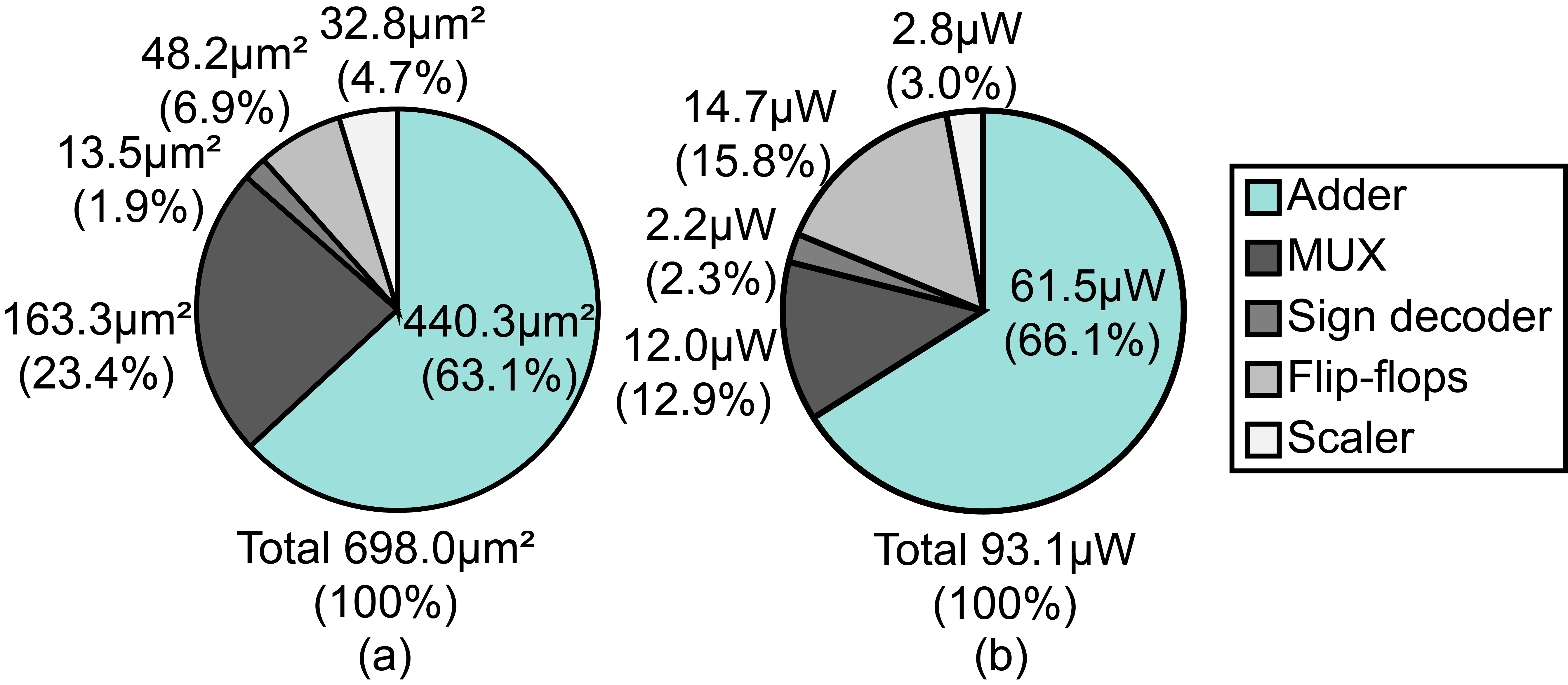}
     \caption{(a) Area and (b) power breakdown of a BS-RAC unit.}
     \label{fig:unit_breakdown}
 \end{figure}

Fig.~\ref{fig:unit_breakdown} shows the area and power breakdown of a BS-RAC unit, which replaces multiplication with selection logic. BS-RAC extends the RAC structure proposed in FIGLUT\cite{park2025figlut} with a lightweight sign decoder to support sign-magnitude BFP operands. The sign decoder contributes only a marginal fraction of the total area and power, indicating that BFP support incurs negligible hardware overhead. 


\subsection{Ablation Study}

 \begin{figure}[t]
     \centering
     \includegraphics[width=0.47\textwidth]{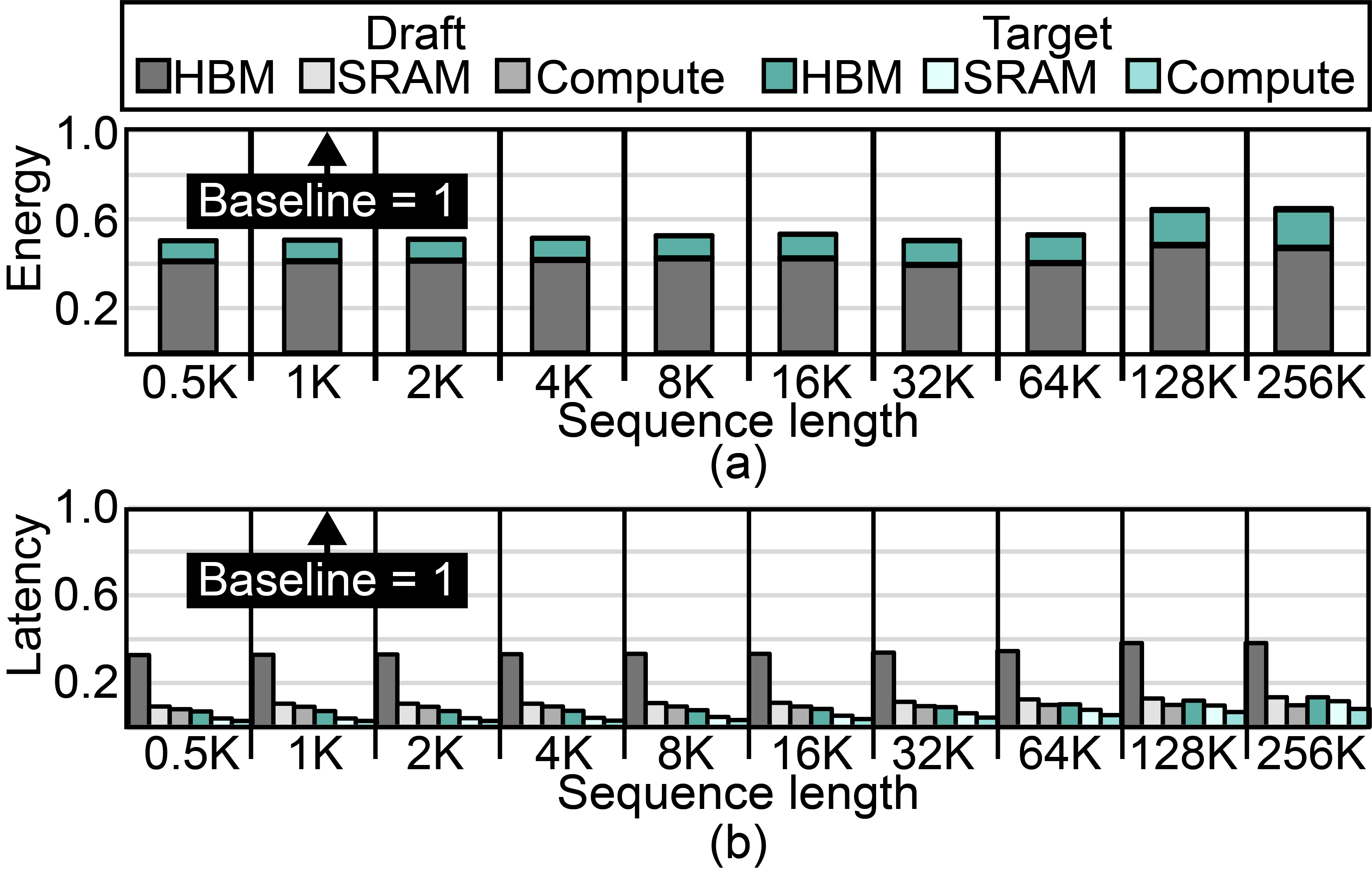}
     \caption{Hardware component breakdown of (a) energy and (b) latency for draft and target execution in BS-Spec on Qwen3-14B with batch size 4. Results are normalized to the autoregressive systolic-array baseline.}
     \label{fig:energy_latency_breakdown}
 \end{figure}

  \begin{figure}[t]
     \centering
     \includegraphics[width=0.49\textwidth]{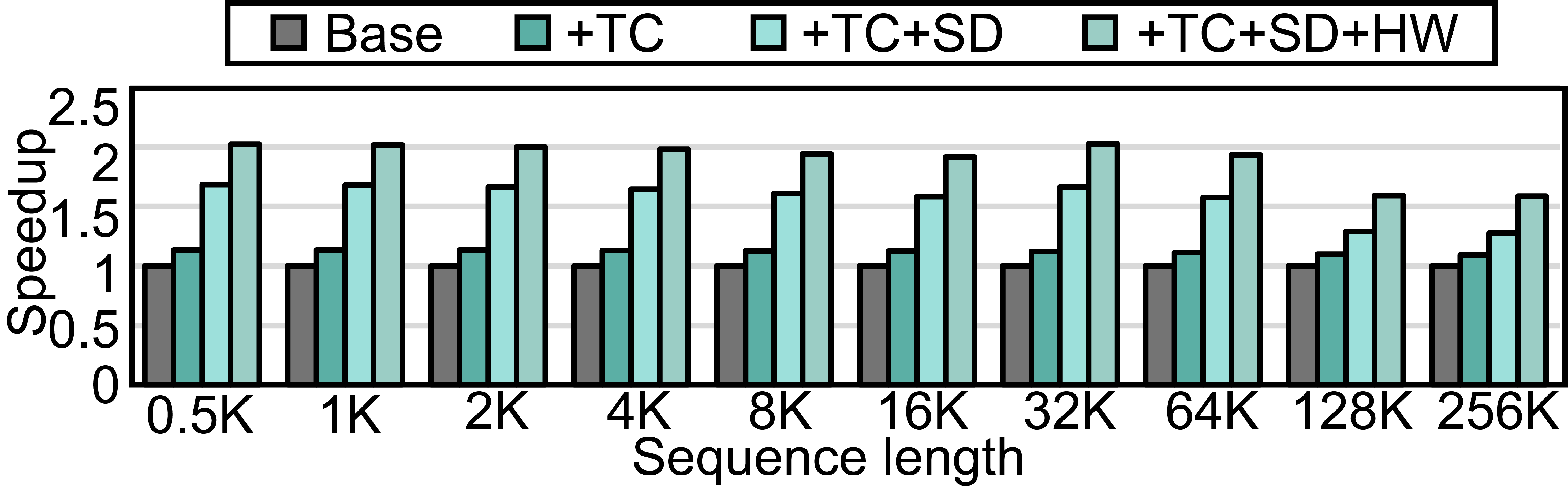}
     \caption{Incremental speedup contributions of target compression (TC), speculative decoding (SD), and dedicated BS-Spec hardware (HW) on Qwen3-14B with batch size 4. Results are normalized to the autoregressive systolic-array baseline.}
     \label{fig:speedup_breakdown}
 \end{figure}

   \begin{figure}[t]
     \centering
     \includegraphics[width=0.49\textwidth]{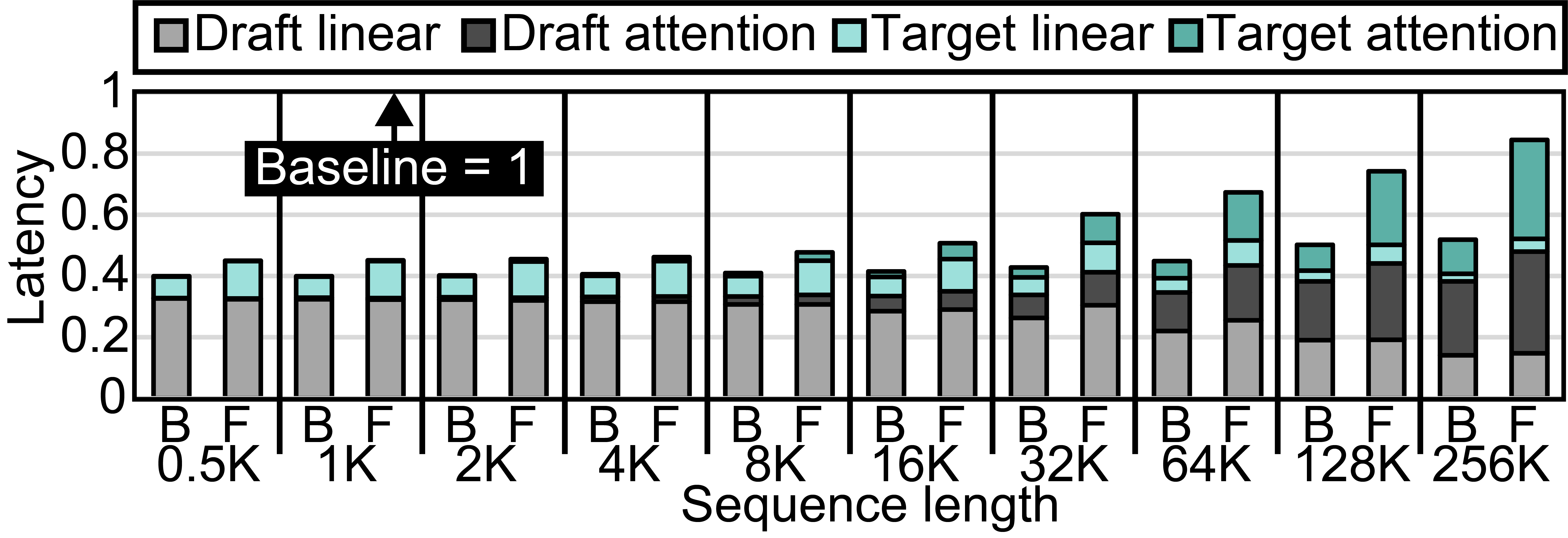}
     \caption{Layer-level latency comparison between the proposed BS-Spec dataflow (B) and the FIGLUT-style dataflow (F) for draft and target linear and attention operations on Qwen3-14B with batch size 4. Results are normalized to the autoregressive systolic-array baseline.}
     \label{fig:stationary_breakdown}
 \end{figure}

\textbf{Hardware Component Breakdown.}
Fig.~\ref{fig:energy_latency_breakdown} presents the component-wise energy and latency breakdown of BS-Spec, including $\gamma$ draft executions and one target execution. In Fig.~\ref{fig:energy_latency_breakdown}(a), energy is accumulated across all executions and hardware components. In Fig.~\ref{fig:energy_latency_breakdown}(b), memory access and computation are pipelined, such that the end-to-end latency is determined primarily by the dominant component together with pipeline overhead. HBM contributes substantially to both draft and target energy and latency, with draft-phase HBM accesses accounting for the largest share overall. Nevertheless, these $\gamma$ draft executions remain considerably cheaper than performing $\gamma$ additional target model decoding steps.

\textbf{Speedup Contribution Breakdown.}
Fig.~\ref{fig:speedup_breakdown} shows the incremental contributions of the main BS-Spec optimizations. Starting from the autoregressive baseline, BFP9 target compression provides a modest gain by reducing target execution cost. Speculative decoding then delivers the largest improvement by replacing multiple target steps with low-cost draft executions and a single verification step. Finally, the proposed LUT-stationary hardware provides additional acceleration over the conventional systolic-array architecture.

\textbf{Dataflow Comparison.}
Fig.~\ref{fig:stationary_breakdown} compares the proposed LUT-stationary hardware with a FIGLUT-style dataflow~\cite{park2025figlut}, both normalized to autoregressive decoding. The proposed design maintains approximately $2\times$ speedup across sequence lengths, whereas the FIGLUT-style mapping degrades as sequence length and target precision increase. This result indicates that the proposed dataflow more efficiently supports both draft and BFP9 target execution.

\section{Conclusion}

This paper presented BS-Spec, a BFP-based bit-sharing framework for self-speculative LLM decoding. BS-Spec derives multiple draft precisions directly from a single target representation through MSB-only partial loading, avoiding separate draft-model storage and additional format conversion. A theoretical offline precision search identifies the weight and KV-cache precision pair that maximizes the estimated speedup for each serving workload, and the resulting configuration is applied at runtime through the PLT. To support this execution efficiently, we developed a bit-serial BFP accelerator with BFP-aware LUT generation and a decoding-oriented LUT-stationary dataflow for variable-precision draft and target computation. Together, these algorithmic and hardware techniques enable efficient workload-adaptive speculative decoding across diverse serving conditions.



\bibliographystyle{IEEEtranS}
\bibliography{refs}

\end{document}